\documentclass[twocolumn]{aastex631}

\usepackage{enumitem}
\usepackage{booktabs}
\usepackage{multirow}
\usepackage{CJK}
\usepackage{bm}
\usepackage{amsmath}
\usepackage{CJK}
\usepackage{natbib}

\defcitealias{Kostov2020}{K20}
\defcitealias{Standing2023}{S23}
\shorttitle{AASTeX v6.3.1 Sample article}
\shortauthors{Wang et al.}

\graphicspath{{./}{figures/}}

\begin{document}
\begin{CJK*}{UTF8}{gbsn}
\title{Photo-dynamical Analysis of Circumbinary Multi-planet system TOI-1338: a Fully Coplanar Configuration with a Puffy Planet}

\author[0000-0003-3015-6455]{Mu-Tian Wang(王牧天)}

\affiliation{School of Astronomy and Space Science, Nanjing University, Nanjing 210023, China.}
\affiliation{Key Laboratory of Modern Astronomy and Astrophysics, Ministry of Education, Nanjing, 210023, People’s Republic of China}
\author[0000-0001-5162-1753]{Hui-Gen Liu(刘慧根)}
\affiliation{School of Astronomy and Space Science, Nanjing University, Nanjing 210023, China.}
\affiliation{Key Laboratory of Modern Astronomy and Astrophysics, Ministry of Education, Nanjing, 210023, People’s Republic of China}

\begin{abstract}

TOI-1338 is the first circumbinary planet system discovered by TESS. It has one transiting planet at P$\sim$95 day and an outer non-transiting planet at P$\sim$215 day complemented by RV observation. Here we present a global photo-dynamical modeling of TOI-1338 system that self-consistently accounts for the mutual gravitational interactions between all known bodies in the system. As a result, the three-dimensional architecture of the system can be established by comparing the model with additional data from TESS Extended Mission and published HARPS/ESPRESSO radial velocity data. We report an inconsistency of binary RV signal between HARPS and ESPRESSO, which could be due to the contamination of the secondary star. According to stability analysis, the RV data via ESPRESSO is preferred. Our results are summarized as follows: (1) the inner transiting planet is extremely coplanar to the binary plane $\Delta I_b \sim 0.12 ^\circ$, making it a permanently transiting circumbinary planet at any nodal precession phases. We updated the future transit ephemerides with improved precisions. (2) The outer planet, despite its untransiting nature, is also coplanar with binary plane by $\Delta I_c=9.1^{+6.0 \circ}_{-4.8}$ (22$^\circ$ for 99\% upper limit). (3) The inner planet could have a density extremely low as $0.137 \pm 0.026$ g/cm$^{-3}$. With a TESS magnitude of 11.45, TOI-1338 b is an optimal circumbinary planet for ground-based follow-up and transit spectroscopy. 
\end{abstract}


\section{Introduction} \label{sec:intro}

Planets orbiting around both of the stars in binary systems are called the circumbinary planets (CBPs). Over the years the detections of CBPs mainly came from transit surveys like Kepler and TESS \citep[e.g., ][]{Doyle2011,Welsh2012,Welsh2015,Kostov2020,Kostov2021}. Subject to the observational bias of the transit method, most confirmed CBPs are coplanar with their host binary orbital planes within $\sim 4.5^\circ$ and close-in (with semi-major axis around 1-2 times of system inner stability limit). Knowing the inclination distribution is essential to understanding the population and occurrence rates of the CBP population \citep{Li2016, Martin2019}. More faraway and misaligned CBPs are harder to detect with transit photometry, as it will produce irregular transit patterns \citep{Martin2014, Chen2021}. However, CBPs elude transiting might dynamically perturb the binary and showcase eclipsing binary variations (ETVs) as potential evidence of their existence \citep[e.g., ][]{Qian2012,Er2021,Esmer2022}, but such claims are sometimes debated \citep{Pulley2022}. Recent radial velocity surveys may expedite the discovery of more misaligned circumbinary planets \citep{Standing2022}. For example, the BEBPOP survey contributes to the discovery of TOI-1338 c \citep[][, hereafter \citetalias{Standing2023}]{Standing2023}.


TOI-1338 system is the second circumbinary multiplanet system around a main-sequence binary, following Kepler-47 \citep{Orosz2012,Orosz2019}. The inner transiting planet TOI-1338 b was discovered by TESS photometry, based on three transits in Sector 3, 6 and 10 \citep[][ hereafter \citetalias{Kostov2020}]{Kostov2020}. Later an outer planet TOI-1338 c at $P\sim215.5$ day is confirmed by radial velocity \citepalias{Standing2023}. Due to the degeneracy of the RV method, the inclination of TOI-1338 c is not known, but the stability analysis in \citetalias{Standing2023} constrains the mutual inclination of TOI-1338 c within $\sim$40$^\circ$, precluding a highly misaligned configuration. TOI-1338 system is a bright ($T=11.45$ mag) CBP system in the southern sky, and TOI-1338 b could have a significant fraction of gas envelope. Therefore, TOI-1338 b is an optimal target for JWST transmission observation, which would provide us with valuable information on the migration history of the planet and, potentially, the atmospheric variability in response to variable incident fluxes. 

After the discovery of TOI-1338 b, TESS revisited TOI-1338 system in the Extended Mission during 2021 and 2023, during which seven more transits of TOI-1338 b were captured. However, the observed transit midtimes offset from the predicted transit ephemerides in \citetalias{Kostov2020}, which assumed a one-planet model, by 0.5-1 transit durations. It could be the result of the perturbation by TOI-1338 c. It's timely to incorporate existing observations to update the future transit ephemerides for the transiting TOI-1338 b.

In this work, we aim to refine the TOI-1338 system parameter. The transit timing/duration variations (TTVs and TDVs) exerted by an external perturbed can help to constrain the three-dimensional architecture of the outer orbit, as has been practiced in some systems around single stars \citep[e.g., ][]{Dawson2014,Mills2017,Almenara2022}. However, the circumbinary planet system could be more complicated, as the TTVs and TDVs are mainly sourced from the inner binary's geometric projection at the time of transit \citep{Armstrong2013,Armstrong2014}. Therefore we perform photodynamical modeling of the TESS light curves and published radial velocity data to self-consistently solve the four-body dynamics. 
The paper is structured as follows. In Section \ref{sec:tess_obs} we briefly introduce the new transit light curves of TOI-1338 b in the TESS Extended Mission used in this work. In Section \ref{sec:photodynam} we detail our photodynamical analysis of the TOI-1338 system, and in Section \ref{sec:photo_result} we present the results and future transit forecast, and conclude in Section \ref{sec:conclusion}.

\section{New Transits of TOI-1338 \lowercase{b} }
\label{sec:tess_obs}

In the following two subsections, we first describe the new TESS observation in Section \ref{sec:tess_lightcurve}. In Section \ref{sec:transit_times} we present the light curve detrending and transit midtimes fitting and compare them with the transit predictions in \citetalias{Kostov2020}.

\subsection{TESS Light Curves \label{sec:tess_lightcurve}}

Ten primary transits of TOI-1338 b are observed during TESS's observation (Sector 3, 6, 10, 27, 30, 34, 37, 62, 65, and 68). The light curve products are made of multiple cadences. We used 30-minute \texttt{gsfc-eleanor-lite} light curves for Sector 3 \citep{Powell_2022}, 2-minute \texttt{SPOC} light curves for S6, 10, 62, 65, and 68, and 20-second \texttt{SPOC} light curves for S27, 30, 34, 37 \citep{Jenkins2016}.

\begin{figure}
    \centering
    \includegraphics[width=0.5\textwidth]{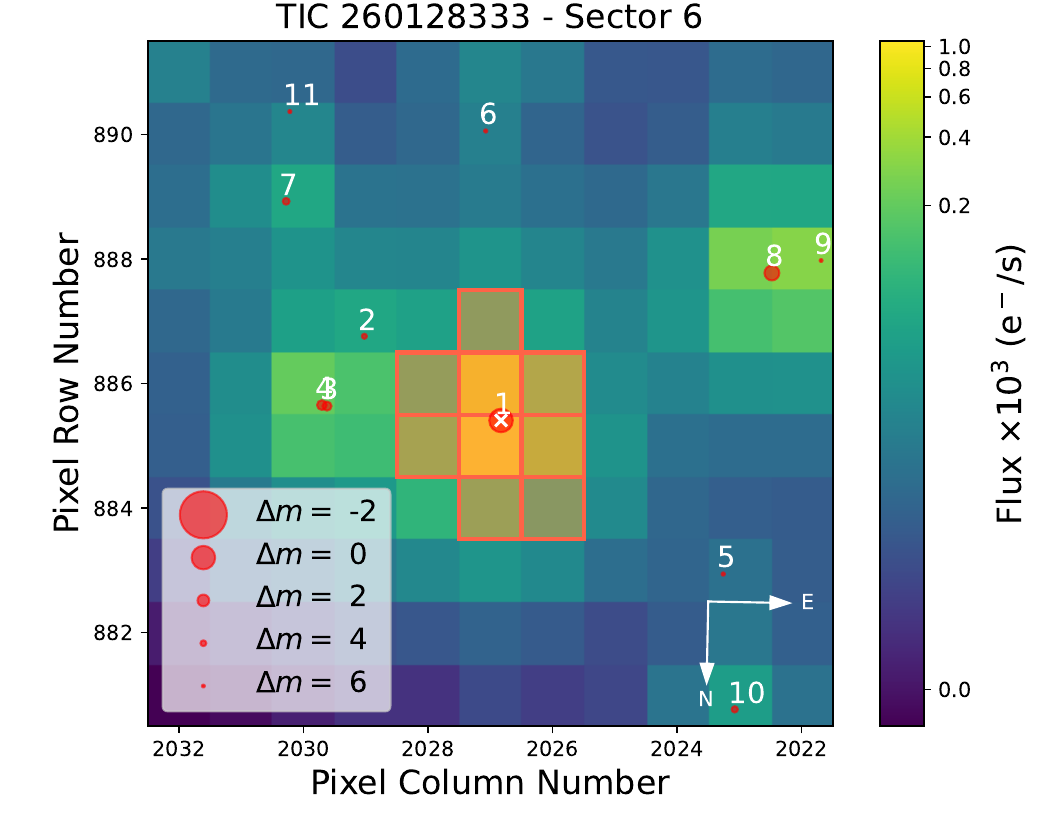}
    \caption{An example of the Target Pixel File of TOI-1338 (the zeroth target) in Sector 6 and the pipeline aperture (orange squares) from \texttt{SPOC}. The three closest nearby sources to TOI-1338 have angular separations of 54.1'', 58.9'', and 60.7'', respectively, all of them have $\Delta G>$ 2.7 mag. This figure is produced by \texttt{tpfplotter} \citep{Aller2020}.}
    \label{fig:tpf}
\end{figure}

The two closest sources to TOI-1338 b have angular separations of 54.1'' and 58.9'', both of them are fainter than TOI-1338 by 3 mags (Figure \ref{fig:tpf}).  None of them fall within the aperture of \texttt{SPOC} in the sectors when CBP transits occur. We also check the depth of the primary eclipses in the sectors of CBP transits from \texttt{SPOC} and GSFC light curves in S3, all of them showing consistent eclipse depth within 1.6\%. The SOAR speckle imaging also indicates the absence of nearby sources within 3'' (see \cite{Tokovinin2019} for a description of instrumentation). Therefore we conclude that the contamination of nearby sources should be minimal in both \texttt{SPOC} and GSFC light curves.

\begin{figure*}[h]
    \centering
    \includegraphics[width=1.0\textwidth]{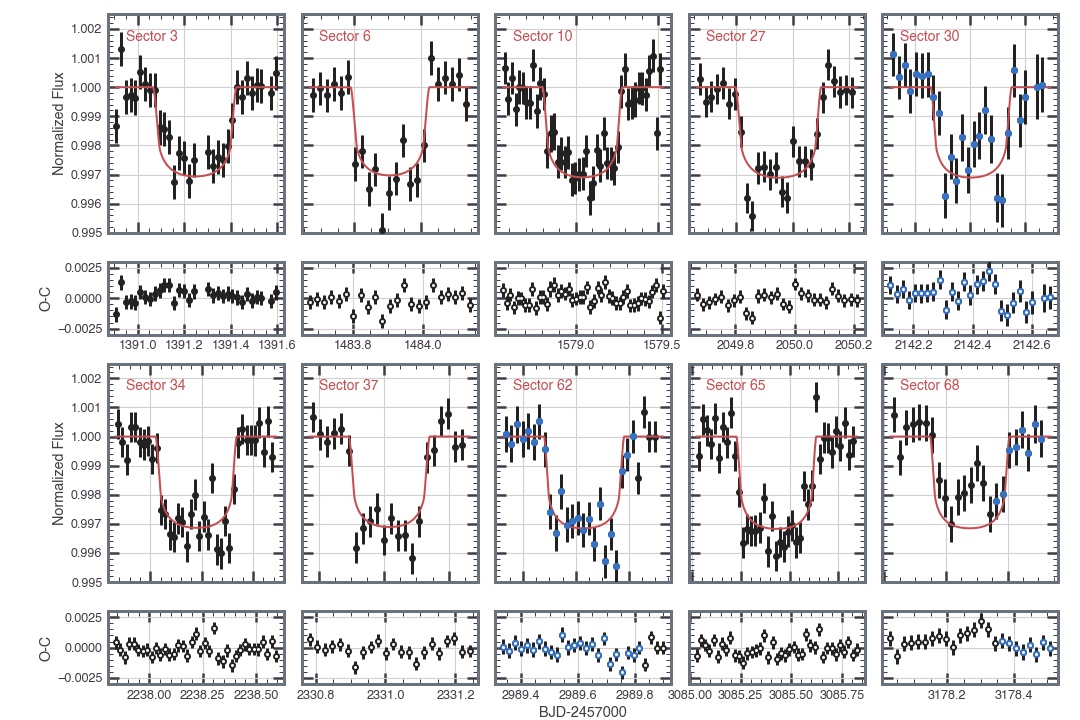}
    \caption{The ten primary transits of TOI-1338 b in TESS observations.  We binned the data into the 30-minute cadence for clarity of display. Red lines are the synthetic light curves from the best-fit photodynamical model of TOI-1338 b (see Section \ref{sec:photodynam}). The residuals between the best-fit model and observation are shown below. Cadences affected by stray light flagged by \texttt{SPOC} are marked as blue. The photodynamical modeling used all primary transits except for the one in Sector 30.}
    \label{fig:best_fit}
\end{figure*}

While most of the transits fall in the window of normal operation, the transit in Sector 30, 62, and 68 suffered from stray light of Earth and Moon (\texttt{SPOC} quality flag= \texttt{1E12}). The stray light causes the background flux to rise and may cause uncertainties in the estimated background fluxes when performing aperture photometry and, thus introducing systematics in the produced light curves. As shown in Figure \ref{fig:best_fit}, the influence of stray light is most pronounced in Sector 30 and 68 light curves: large photometric jitters are present around the mid-transit and the post-transit portion of light curves. The transit in Sector 62 has a fall-off in fluxes near the egress phase. When binned to a 30-minute cadence, the egress/ingress of CBP transit is still well-resolved in Sector 62 and 68, but the egress is hardly resolved in Sector 30.  

The variations in transit duration and midtimes are unique features for transiting CBP, and resolving the ingress/egress phase is crucial to precisely determining the magnitude of variations and measuring the movement and coordinates of different bodies. We decided to fit all the transits except for the one in Sector 30 due to its damaged egress phase. 
Later in Section \ref{sec:systematics} we will show the in-transit systematics will indeed affect the transit midtime determinations if we directly fit the light curves, but the bias is less severe if we fit the light curves globally using photodynamical modeling.

\subsection{Detrending and Transit Times Fitting \label{sec:transit_times}}

We used the \texttt{sap\_flux} product to measure the midtime and duration for each transit. The detrending and fitting are devised as follows. 
The light curves are first fitted with quadratic limb darkening transit model in \texttt{batman} \citep{Kreidberg2015}.  The fitted parameters are transit midtime $T_{\rm mid}$, planet period $P$, semi-major axis scaled to stellar radius $a/R_*$, impact parameter $b$, planetary-to-stellar radii ratio $R_p/R_*$, and two quadratic limb darkening coefficients $q_1,~q_2$ \citep{Kipping2013}. We assume circular orbits during the midtime fitting. The $\chi^2$ likelihood function is used to measure the goodness-of-fit of the model compared to the observation. The model is first optimized using Monte Carlo Markov Chain (MCMC) for 30000 iterations. Once a best-fit model is found, the light curves are trimmed to shorter segments centered at the best-fit midtime, with lengths of two times of best-fit transit durations. The trimmed light curves are detrended with a linear slope, with in-transit data given zero weight during the linear fit. The flux errors are also rescaled to make the reduced $\chi^2=1$. The scaling factors for transits in each sector are also listed in Table \ref{tab:transit_time_duration}. Then the model is re-optimized using MCMC for another 30000 iterations.  We also calculated the estimated transit duration from our fitted parameters via Equation 14\footnote{To the first order, we assume the binary-planet relative velocity is constant during the CBP transits, therefore the calculation of transit durations can be approximated to be similar to transits around the single-star system. However, the binary-planet relative velocity is constantly changing throughout the transits due to the lateral motion of binary, thus the transit profiles could be asymmetric in ingress/egress for circumbinary planets \citep[see Figure 3 in][]{Liu2013}. A more precise transit duration estimate for transiting circumbinary planets would be derived from a photodynamical model. For TOI-1338 b, the asymmetry in transit ingress/egress is at most 15 minutes.} in \cite{winn10}.

The measured transit midtimes and durations are listed in Table \ref{tab:transit_time_duration}.
Comparing the predicted transit mid-times of \citetalias{Kostov2020} and the observed mid-times, we found that while the observation in Sector 37 matches well with the prediction in \citetalias{Kostov2020}, the rest of the observed transit midtimes deviates from \citetalias{Kostov2020} predictions by more than the magnitude of their predicted uncertainties, which is most evident in Sector 65 and 68 where deviations are readily comparable to the transit durations. 
    The deviation between \citetalias{Kostov2020} predictions and TESS following observations could be caused by the imprecise system parameters, most likely the planetary orbital parameters, derived from the limited three transits observed in the TESS Primary Mission used in \citetalias{Kostov2020}, 
    and also could source from the gravitational interactions from the TOI-1338 c whose presence is unknown at that time. 
In Section \ref{sec:mutual_inclination} we will show that TOI-1338 c is indeed perturbing TOI-1338 b and contributes to the transit timing deviations we show here. Therefore we carry out photodynamical modeling to update the global parameters of TOI-1338 system in the next section.


\begin{deluxetable*}{cccccc}
\label{tab:transit_time_duration}
\tablecaption{Comparison Between the Predicted and Observed Transit Midtimes and Durations of TOI-1338 b}
\tablehead{\colhead{Sector} & \colhead{Transit Midtime} & \colhead{Duration} & \colhead{Error inflation factor} & \colhead{Predicted Midtime\tablenotemark{a}}\\ 
\colhead{} & \colhead{(BJD-2457000)}  & \colhead{(hr)} & \colhead{} & \colhead{}} 
\startdata
3 	&	 1391.2516 $\pm$ 0.0073  &	 8.5584 	$\pm$	0.4226 	&	 0.7539 	&   -                     \\
6 	&	 1483.9048 $\pm$ 0.0029	 &	 5.9564     $\pm$   0.2135 	&	 1.0100 	&   -                     \\
10 	&	 1579.0423 $\pm$ 0.0052	 &	 11.5838    $\pm$   0.4883  &	 1.0110 	&   -                     \\
27 	&	 2049.9515 $\pm$ 0.0034	 &	 7.1834     $\pm$   0.3704 	&	 1.0309 	&   2049.6930$\pm$ 0.0196 \\
30 	&	 -  	                 &	 			 -              &	 - 	        &   2142.3991$\pm$ 0.0224 \\
34 	&	 2238.2258 $\pm$ 0.0034	 &	 9.5885     $\pm$ 0.2923 	&	 1.0162 	&   2237.9229$\pm$ 0.0236 \\
37 	&	 2331.0148 $\pm$ 0.0024	 &	 5.8707     $\pm$ 0.1752 	&	 0.9953 	&   2330.9984$\pm$ 0.0264 \\
62 	&	 2989.6356 $\pm$ 0.0047	 &	 7.6709     $\pm$ 0.4334 	&	 0.9886 	&   2989.3540$\pm$ 0.0509 \\
65 	&	 3085.4243 $\pm$ 0.0044	 &	 10.2463    $\pm$ 0.3505    &	 0.9953 	&   3084.8165$\pm$ 0.0693 \\
68 	&	 3178.2853 $\pm$ 0.0077	 &	 6.2203     $\pm$ 0.6371    &	 0.9498 	&   3177.9790$\pm$ 0.0600 \\
\enddata
\tablenotetext{a}{Prediction from \citetalias{Kostov2020}}
\end{deluxetable*}

\section{Photodynamical Modeling \label{sec:photodynam}}
In this section, we present the photodynamical analysis of TOI-1338. We firstly describe our photodynamical model in Section \ref{sec:model_setup} and the system parameter retrieval strategy in Section \ref{sec:sampling}; we also discuss some potential bias by transit systematics in Section \ref{sec:systematics} and test the validity of photodynamical solutions in terms of the dynamical stability in Section \ref{sec:stability}.

\subsection{Model Setup \label{sec:model_setup}}

The system parameters of circumbinary planet systems are often exhaustively explored by photodynamical modeling (e.g., \cite{Doyle2011a}, \cite{Welsh2012}). Given initial orbital and mass parameters, the N-body integrator solves the equations of motions of binary and planets and produces in-time observables, i.e., eclipse, transit light curves, and radial velocity time series. The synthetic data are compared to the observations, and system parameter credible intervals are explored by sampling algorithms (e.g., MCMC) with related observational uncertainties.
 
The dynamical status of a two-planet circumbinary system can be characterized by 22 parameters, equivalent to six binary and twelve planetary osculating orbital elements ($P,~e,~i,~w$, $\Omega$ and mean longitude $\lambda$), and four masses ($M_A,~M_B,~M_b,~M_c$). The ascending node angle of the binary orbit can be further set to zero, which leaves 21 parameters. 
Synthetic stellar eclipse and planet transit light curves can be produced by the relative positions of each body, which require the stellar and planetary radii ($R_A,~R_B,~R_b$), primary-to-secondary star surface brightness ratio in TESS band $f_{\rm TESS}$, and two sets of quadratic law limb darkening coefficients, for which we used the triangular sampling in \cite{Kipping2013}. In total 6 additional parameters are needed for light curve synthesis.
The secondary eclipses show flat-bottom in-eclipse profiles and have low SNR, therefore the limb darkening coefficients of the secondary star are not well constrained and thus are set to be the same as the primary star\footnote{A more sensible treatment for the limb darkening coefficients of the secondary star would be using the theoretical values presented by \cite{Claret2017}: $u_1=0.15,u_2=0.47$ for $T_{\rm eff}=3300$ K, $\log g$=5.0, and [Fe/H]=0.0. Using these theoretical values, the maximum difference between the new secondary eclipse light curves and our adopted model is 10 ppm, only manifesting during the ingress/egress phase. The error is much smaller than the observational error of 200 ppm and thus this barely changes the overall likelihood of the photodynamical fitting and parameter sampling.}.
The dynamical modeling also issues radial velocity curves relative to the system barycenter. Thus, the system barycentric velocity $\gamma$ is modeled as three instrument-dependent parameters for HARPS and ESPRESSO(pre-2019 and post-2019). 
In summary, 30 parameters are needed in photo-dynamical modeling in the case of the TOI-1338 system.

We outline our photodynamical model as follows. The system is initialized in the Jacobian coordinate, namely, the object is initialized in the order of its semi-major axis to the primary star. 
We use the \texttt{IAS15} integrator in the python package \texttt{rebound} \citep{Rein&Liu2012} to solve the motion of all objects. 
The integration is done in the center-of-mass coordinate. 
We use \texttt{batman} \citep{Kreidberg2015} to generate synthetic light curves from the relative coordinates of different bodies on the sky-projected plane centered on the system barycenter, after linear corrections for the light travel effect.
The radial velocity data precision of ESPRESSO achieved $\sim$1 m/s, calling the need for relativistic correction, therefore the simulated RVs are further corrected with light travel effects, transverse Doppler, and gravitational reddening effects \citep{Zucker2009,Konacki2010,Sybilski2013}. However, later we found that these will not significantly alter the fitted result as the RV variations are absorbed into the barycenter velocity. Tidal effects are not included since it suffered from uncertainty in tidal parameters and is estimated to have an amplitude $\sim 0.2$ m/s, which is much lower than the total amplitude of relativistic effects ($\sim 5$ m/s).

For the case of the TOI-1338 system, the General Relativity (GR)-induced apsidal precession is around 9.01 arcsec per year. The apsidal precession rate induced by tides is $>10$ times smaller than the GR term, following the prescription in \cite{correia2013}. We do not include the GR and tidal effects in our dynamical model.

The reference epoch is fixed at BJD=2458300.00, approximately 90 days before the first observed CBP transit in Sector 3. Using the best-fit values in \citetalias{Kostov2020}, we integrate the TOI-1338 system into our reference epoch and record the result as an initial guess of the model parameters. The orbital parameters of the outer planet are derived from \citetalias{Standing2023}, with randomly generated node angles and inclinations but satisfying the system stability limit (mutual inclination smaller than 40$^\circ$, see Supplement Figure 8 of \citetalias{Standing2023}). 

We adopt the differential evolution Monte Carlo Markov Chain \citep[DE-MCMC,][]{TerBraak2006} in \texttt{emcee} to sample the parameter space. The range of parameters is displayed in Table \ref{tab:model_bestfit}. Flat priors are assumed for all parameters. The $\chi^2$ function is used to define the goodness-of-fit of the model. This is computed by the square of the differences between observation and synthetic photometric/RV data, divided by the square of observational errors. 

\subsection{Fitted Data}

Using the photodynamical model described in the above section, we simultaneously fitted the binary eclipse light curves, transits of TOI-1338 b in TESS light curves, and the radial velocity data published in \citetalias{Standing2023}.

We fitted nine TOI-1338 b's primary transit events, excluding one transit event in Sector 30 since it is severely affected by systematics. To speed up the integration speed and prevent the potential bias delivered by the instrumental systematics or out-of-eclipse trend \citep{Orosz2019}, we select nine primary and secondary eclipses, respectively, for the following photodynamical modeling, based on the ranking of their reduced $\chi^2$. Figure \ref{fig:primary_eclipse_fit} and \ref{fig:secondary_eclipse_fit} in Appendix \ref{sec:pri_sec_eclipse_profile} show the selected nine primary eclipses and secondary eclipses. The transit and eclipse data are detrended according to Section \ref{sec:transit_times}. Once an initial dynamical result that can reproduce the observed light curves is obtained, the light curves are detrended again with the  more precise transit/eclipse midtimes and durations inferred from the dynamical result.

61 HARPS and 123 ESPRESSO RV data points are published by \citetalias{Standing2023}. Many of these RV data have been flagged to be potential FWHM/bisector outliers, or due to the data being taken at the time of binary or planetary transits. We exclude RV data with outlier flags for our photodynamical model, leaving 58 HARPS and 103 ESPRESSO RV data to be modeled.

\subsection{Planetary System Parameters Sampling \label{sec:sampling}}

We first perform a \textit{joint} fit to the stellar eclipse, transit photometry, and HARPS/ESPRESSO radial velocity data.
At the starting stage, 100 chains evolved from the initial guess with small offsets to explore the parameter space. 
After the chains remained stable in small ranges, we evolved 100 chains for another 690000 steps. We remove the first 350000 steps as the burn-in phase. The rest of the 340000 steps from 100 chains are concatenated together, and they are sampled every 1000 steps to make the posterior sample with a total size of 34000. The Gelman-Rubin statistics $\hat{R}$ of all parameters are smaller than 1.05 \citep{Brooks1998}.  The autocorrelation steps of fitted parameters vary between 1972 and 5213, giving at least 6522 effectively independent samples in total. The median and uncertainties (defined as the 16th and 84th percentiles of the posterior distribution) of each fitted parameter are listed in Table \ref{tab:model_bestfit}.

We examined the residuals of all eclipse, transit, and RV residuals and found the best-fit solution from \textit{joint} fit provides a good match to the light curve data. However, we found a very significant periodicity corresponding to the binary period in HARPS RV residuals (FAP $>$ 0.1\%) but none in ESPRESSO residuals. In other words, there is inconsistency in the binary RV motion between HARPS and ESPRESSO RV data. The peak-to-peak amplitude of HARPS residual is $\sim 14$ m/s, and peaks around when the binary is in its conjunction. This is not likely a problem associated with our photodynamical code as this inconsistency persists when we perform a joint RV-only Keplerian fit analysis (see details in Appendix \ref{sec:app_harps_residuals}). 

To circumvent the potential bias on the binary orbital parameters brought by RV inconsistency, we experimented with two additional photodynamical fits performed individually to ESPRESSO and HARPS data (and hereafter we refer to these two experiments as \textit{esp}-only and \textit{harps}-only analyses), with the light curve dataset identical to the \textit{joint} fit. We took 100 samples from the \textit{joint} fit posterior samples and evolved them for 300000 steps.  We trimmed the first 150000 steps as the burn-in phase, and the Gelman-Rubin statistics $\hat{R}$ for the remaining chains are smaller than 1.04 for all parameters. The effective sample sizes are at least 4263 and 3624 for \textit{esp}-only and \textit{harps}-only posteriors, respectively. We kept the last 150000 steps and sampled every 1000 steps to make the posterior distributions. The median and uncertainties of fitted parameters from the \textit{esp}-only and \textit{harps}-only analysis are presented in the second and third columns of Table \ref{tab:model_bestfit}. 

Most parameters from \textit{esp}-only and \textit{harps}-only fit are 1-2 $\sigma$ deviant from \textit{joint} analysis, while some of them show 3$\sigma$ discrepancy.
    For the binary orbital parameters, the \textit{joint} fit and \textit{esp}-only fit show close agreement, while the binary eccentricity from \textit{harps}-only fit is around 3-$\sigma$ larger than those from the other two analyses, which probably stems from the RV inconsistency from HARPS data. 
    The binary mass ratio and orbital parameters of TOI-1338 b inferred from the three analyses are different too, with \textit{joint} and \textit{esp}-only analysis having 3$\sigma$ discrepancy on binary mass ratio and that from \textit{harps}-only being intermediate between the former two. 
We provide some intuitions of the potential causes. 
TOI-1338 is a single-line binary, so only the mass function (related to the mass ratio, absolute mass, and binary period, eccentricity, inclination) can be constrained from RV data. The degeneracy between the mass ratio and the mass function is solved by accounting for the exact timing of planetary transit. 
    For example, the peak-to-peak amplitude of transit timing variations nonlinearity of CBP transit is, in the first order, related to the binary phase and the exact projected stellar location, which is further related to the semi-major axis of the transited star and hence the mass ratio can be constrained \citep{Schwamb2013,Armstrong2013,Kostov2013,Kostov2014}. 
    The CBP transit durations also depend on the relative velocity difference between the binary star and planet. For the latter one, the velocity of the planet is also related to the total masses of the inner binary, and thus can also contribute to solving the mass ratio from the mass function. 
    This degeneracy is also supported by the clear correlation between the mass ratio and the inner planetary orbit parameters ($P_b$, $\lambda_b$, eccentricity vectors).
Following this thread of thought, we attributed the discrepancy of mass ratio derived from the two analyses to the difference in planetary orbit solutions, since we found the derived mass function from \textit{joint} and \textit{esp}-only analysis is consistent within 1$\sigma$.

Despite the difference in inner planetary orbit solutions, the best-fit $\chi^2$ of the transit light curve between \textit{joint} and \textit{esp}-only analysis has a negligible difference, which is $\chi^2_{\rm transit} = 10166.8$ and $\chi^2_{\rm transit} = 10164.6$ for 10183 cadences, respectively. But given the inconsistency between HARPS and ESPRESSO data, we prefer \textit{esp}-only solution over the \textit{joint}-fit one.
Between \textit{esp}-only and \textit{harps}-only solutions, the $\chi^2$ in transit is almost identical, and the \textit{harps}-only solution is only marginally preferred over the \textit{esp}-only ones in terms of the $\chi^2_{\rm eclipse}$. However, we think the outer planet's orbit is better constrained by the ESPRESSO data, for its higher precision and more observations. 

At the writing of the manuscript, \cite{Sebastian2024} published the absolute dynamical masses of TOI-1338 AB for $1.098\pm0.017$ M$_\odot$ and $0.307 \pm 0.003$ M$_\odot$, respectively. This measurement is more aligned with our \textit{esp}-only solution. Additionally, in Section \ref{sec:stability} we will show that the \textit{esp}-only solution are more dynamically stable compared to the other two sets of solutions. Therefore, among the three analyses presented in this section, we adopt the solution of \textit{esp}-only analysis. 


In our nominal photodynamical model, we assume constant zero  flux dilution for all TESS sectors. For completeness of the photodynamical discussion, we also run a model accounting for different dilution levels in different sectors. Applying a unique dilution factor to each sector is necessary for missions like TESS since the star's location in the CCD plane in each sector is different and so is the amount of flux from the nearby stars expected to contain within the target's aperture. In this simulation, we include the primary and secondary eclipse light curves in the same nine sectors as the transits were observed to help better constrain the dilution factor. The dilution factors are allowed for negative values to reflect the situation where the background fluxes are potentially over-substracted. However, we found the dilution factor of each sector is consistent with zero within 1$\sigma$, and the important parameters like stellar mass ratios did not change significantly. We attribute this to nearly consistent or negligible dilution within TOI-1338 in different sectors and proceed with the remaining discussion with the model without dilution consideration.

\begin{deluxetable*}{lccccc}
\label{tab:model_bestfit}
\tablecaption{Priors, posterior median and 68.3\% (1$\sigma$) credible intervals of physical parameters from the photodynamical model.}
\tablehead{\colhead{Paramter}& \colhead{Unit} & \colhead{Prior Range} & \colhead{TESS+ESP+HARPS} & \colhead{\textbf{TESS+ESP (adopted)}} & \colhead{TES+HARPS}} 
\startdata  
\textbf{\underline{Binary}} &&&& \\
$M_A$       &   M$_\odot$   &   [0.8,1.3]        & 	 $1.1489^{+0.0075}_{-0.0076}$ 	 		& 	 $1.0936^{+0.0072}_{-0.0072}$ 	     & 	 $1.1224^{+0.0095}_{-0.0097}$ 	 \\ 
$q$         &   -           &   [0.26,0.34]      & 	 $0.2753^{+0.00071}_{-0.00069}$ 	   	& 	 $0.28065^{+0.00073}_{-0.00072}$     & 	 $0.27783^{+0.00094}_{-0.00091}$ 	 \\ 
$R_A$       &   R$_\odot$   &   [1.0,1.5]        & 	 $1.3334^{+0.0038}_{-0.0038}$ 		    & 	 $1.313^{+0.0038}_{-0.0038}$ 	     & 	 $1.3246^{+0.0043}_{-0.0044}$ 	 \\ 
$R_B$       &   R$_\odot$   &   [0.2,0.4]        & 	 $0.31063^{+0.0009}_{-0.00091}$ 	 	& 	 $0.30582^{+0.00094}_{-0.00094}$ 	 & 	 $0.3086^{+0.0011}_{-0.0011}$ 	 \\ 
$P_{B}$   &   day         &   [14.6,14.7]      & 	 $14.6085738^{+3.4e-06}_{-3.5e-06}$ 	& 	 $14.6085659^{+6.2e-06}_{-5.7e-06}$  & 	 $14.608572^{+5.5e-06}_{-4.9e-06}$ 	 \\ 
$e_{B}$   &   -           &   [0.15,0.16]      & 	 $0.155498^{+1e-05}_{-1e-05}$ 	        & 	 $0.155489^{+1.1e-05}_{-1e-05}$ 	 & 	 $0.155633^{+3e-05}_{-3e-05}$ 	 \\ 
$i_{B}$   &   deg         &   [89,92]          & 	 $90.415^{+0.042}_{-0.044}$ 	        & 	 $90.403^{+0.045}_{-0.047}$ 	     & 	 $90.426^{+0.045}_{-0.047}$ 	 \\ 
$\omega_{B}$& deg         &   [80,150]         & 	 $117.7692^{+0.0038}_{-0.0038}$ 	    & 	 $117.7638^{+0.0042}_{-0.0041}$ 	 & 	 $117.7885^{+0.0095}_{-0.0096}$ 	 \\ 
$\lambda_{B}$&   deg      &   [225,310]        & 	 $270.1003^{+0.0015}_{-0.0016}$ 	 & 	 $270.0958^{+0.0017}_{-0.0017}$ 	 & 	 $270.1157^{+0.0025}_{-0.0025}$ 	 \\ 
$f_{\rm TESS}$ &       -    &   [0.99,1.0]       & 	 $0.995856^{+3.8e-05}_{-3.8e-05}$ 	    & 	 $0.995856^{+3.8e-05}_{-3.8e-05}$ 	 & 	 $0.995857^{+3.8e-05}_{-3.8e-05}$ 	 \\ 
$q_{\rm 1,A}$  &       -    &   [0,1]            & 	 $0.169^{+0.022}_{-0.021}$ 	            & 	 $0.17^{+0.021}_{-0.021}$ 	         & 	 $0.169^{+0.022}_{-0.02}$ 	 \\ 
$q_{\rm 2,A}$  &       -    &   [0,1]            & 	 $0.479^{+0.052}_{-0.049}$ 	            & 	 $0.478^{+0.051}_{-0.047}$ 	         & 	 $0.48^{+0.051}_{-0.048}$ 	 \\ 
\hline    
\textbf{\underline{Planet b}}&&&& \\
$P_b$       &   day         &   [94.5,96.5]      & 	 $95.4223^{+0.0094}_{-0.0099}$ 	 & 	 $95.4001^{+0.0062}_{-0.0056}$ 	 & 	 $95.3995^{+0.0088}_{-0.0097}$ 	 \\ 
$i_b$       &   deg         &   [89.0,91.5]      & 	 $90.494^{+0.013}_{-0.013}$ 	 & 	 $90.494^{+0.013}_{-0.014}$ 	 & 	 $90.497^{+0.013}_{-0.013}$ 	 \\ 
$\lambda_b$ &   deg         &   [80,140]         & 	 $104.39^{+0.2}_{-0.22}$ 	     & 	 $103.28^{+0.19}_{-0.19}$ 	     & 	 $103.36^{+0.26}_{-0.27}$ 	 \\ 
$\Omega_b$  &   deg         &   [-0.5,0.5]       & 	 $-0.046^{+0.034}_{-0.034}$ 	 & 	 $-0.076^{+0.036}_{-0.036}$ 	 & 	 $-0.052^{+0.036}_{-0.036}$ 	 \\ 
$R_b$       &R$_{\rm Jup}$  &   [0.5,0.8]        & 	 $0.6958^{+0.0049}_{-0.0049}$ 	 & 	 $0.6835^{+0.0047}_{-0.0047}$ 	 & 	 $0.6912^{+0.0055}_{-0.0054}$ 	 \\ 
$\sqrt{e_b}\cos\omega_b$&-  &   [-0.5,0.5]       & 	 $0.002^{+0.012}_{-0.012}$ 	     & 	 $0.0564^{+0.0074}_{-0.0074}$ 	 & 	 $0.051^{+0.01}_{-0.011}$ 	 \\ 
$\sqrt{e_b}\sin\omega_b$&-  &   [-0.5,0.5]       & 	 $0.1488^{+0.0075}_{-0.0078}$ 	 & 	 $0.1729^{+0.0039}_{-0.004}$ 	 & 	 $0.17^{+0.0059}_{-0.0058}$ 	 \\ 
$M_b$       &M$_{\rm Jup}$  &   [0.0,0.1]        & 	 $0.0204^{+0.0059}_{-0.0059}$ 	 & 	 $0.0355^{+0.0066}_{-0.0067}$ 	 & 	 $0.0104^{+0.0088}_{-0.0067}$ 	 \\ 
\hline
\textbf{\underline{Planet c}} &&&& \\
$P_c$       &   day         &   [200,235]        & 	 $211.67^{+0.64}_{-0.64}$ 	 & 	 $215.79^{+0.46}_{-0.51}$ 	 & 	 $210.25^{+0.57}_{-0.51}$ 	 \\ 
$i_c$       &   deg         &   [50,130]         & 	 $95.5^{+3.6}_{-3.9}$ 	     & 	 $97.0^{+6.7}_{-6.8}$ 	     & 	 $95.2^{+5.3}_{-5.5}$ 	 \\ 
$\lambda_c$ &   deg         &   [0,360]          & 	 $116.8^{+6.2}_{-6.1}$ 	     & 	 $142.8^{+5.4}_{-6.0}$ 	     & 	 $119.6^{+6.6}_{-6.6}$ 	 \\ 
$\Omega_c$  &   deg         &   [-50,50]         & 	 $9.5^{+3.8}_{-4.4}$ 	     & 	 $3.4^{+4.5}_{-5.0}$ 	     & 	 $12.7^{+5.6}_{-6.1}$ 	 \\ 
$\sqrt{e_c}\cos\omega_c$&-  &   [-0.55,0.55]     & 	 $0.25^{+0.04}_{-0.04}$ 	 & 	 $0.17^{+0.08}_{-0.126}$ 	 & 	 $0.312^{+0.034}_{-0.042}$ 	 \\ 
$\sqrt{e_c}\sin\omega_c$&-  &   [-0.55,0.55]     & 	 $-0.236^{+0.04}_{-0.036}$ 	 & 	 $-0.046^{+0.072}_{-0.068}$  & 	 $-0.052^{+0.075}_{-0.074}$ 	 \\ 
$M_c\sin i_c$&M$_{\rm Jup}$ &   [0.0,0.6]        & 	 $0.2018^{+0.0084}_{-0.0082}$& 	 $0.234^{+0.01}_{-0.01}$ 	 & 	 $0.221^{+0.018}_{-0.018}$ 	 \\ 
\hline
\textbf{\underline{Instrument Parameters}}&&&& \\
$\gamma_{\rm HARPS}$ &  km/s    &   [30.5,31.0]  & 	 $30.76491^{+0.00048}_{-0.00048}$ 	 & 	 -                                 	 & 	 $30.76524^{+0.00049}_{-0.0005}$ 	 \\ 
$\gamma_{\rm ESP19}$ &  km/s    &   [30.5,31.0]  & 	 $30.61922^{+0.00045}_{-0.00045}$ 	 & 	 $30.61812^{+0.00045}_{-0.00047}$ 	 & 	 -                                 	 \\ 
$\gamma_{\rm ESP21}$ &  km/s    &   [30.5,31.0]  & 	 $30.61569^{+0.00018}_{-0.00018}$ 	 & 	 $30.61536^{+0.00018}_{-0.00018}$ 	 & 	 - 	                                 \\ 
\hline
\textbf{\underline{Derived Parameters}}&&&& \\
$M_B$		&	M$_\odot$	&		-		 & 	 $0.3163^{+0.0013}_{-0.0013}$ 	 	& 	 $0.3069^{+0.0012}_{-0.0012}$ 	 		  & 	 $0.3118^{+0.0016}_{-0.0016}$ 	 \\
$M_b$       &   M$_\oplus$  &       -        & 	 $6.5^{+1.9}_{-1.9}$ 	            & 	 $11.3^{+2.1}_{-2.1}$ 	                  & 	 $3.3^{+2.8}_{-2.1}$ 	 \\ 
$M_c$       &   M$_\oplus$  &       -        &   $64.6^{+2.8}_{-2.7}$ 	            & 	 $75.4^{+4.0}_{-3.6}$ 	                  & 	 $70.8^{+6.3}_{-5.8}$ 	 \\ 
$R_b$       &   R$_\oplus$  &       -        & 	 $7.799^{+0.055}_{-0.055}$ 	        & 	 $7.661^{+0.053}_{-0.053}$ 	              & 	 $7.748^{+0.062}_{-0.061}$ 	 \\ 
$e_b$       &   deg         &       -        & 	 $0.0223^{+0.0023}_{-0.0023}$ 	 	& 	 $0.0331^{+0.0022}_{-0.0021}$  	 		  & 	 $0.0315^{+0.0031}_{-0.0028}$ 	 \\ 
$e_c$       &   deg         &       -        & 	 $0.12^{+0.014}_{-0.013}$ 	 		& 	 $0.037^{+0.032}_{-0.026}$($<0.086$,95\%) & 	 $0.105^{+0.02}_{-0.021}$ 	 \\ 
$\omega_b$  &   deg         &       -        & 	 $89.4^{+4.8}_{-4.6}$ 	 			& 	 $71.9^{+1.9}_{-1.8}$ 	 				  & 	 $73.3^{+3.0}_{-2.7}$ 	 \\ 
$\omega_c$  &   deg         &       -        & 	 $316.6^{+8.8}_{-8.4}$ 	 			& 	 $344.64^{+25.87}_{-31.25}$ 	 		  & 	 $350.68^{+13.46}_{-14.25}$ 	 \\ 
$\Delta I_b$&   deg         &       -        & 	 $0.099^{+0.027}_{-0.023}$ 	 		& 	 $0.127^{+0.025}_{-0.024}$ 	 			  & 	 $0.098^{+0.028}_{-0.024}$ 	 \\ 
$\Delta I_c$&   deg         &       -        & 	 $11.5^{+3.4}_{-3.7}$ 	 			& 	 $9.1^{+6.0}_{-4.8}$ 	 				  & 	 $14.4^{+5.7}_{-5.7}$ 	 \\ 
\hline
\textbf{\underline{Best-fit $\chi^2$}}&&&& \\
$\chi^2_{\rm total}$      &&& 15768.7 & 15650.7 & 15386.7 \\
$\chi^2_{\rm transit}$    &&& 10164.7 & 10166.86 & 10164.4 \\
$\chi^2_{\rm eclipse}$    &&& 5166.36 & 5169.92 & 5164.72 \\
$\chi^2_{\rm RV}$         &&& 437.82 & 304.96 & 58.06 \\
\enddata
\tablecomments{Osculating jacobian orbital parameters valid at BJD=2458300.0}
\end{deluxetable*}

\subsection{The Effect of In-Transit Noise \label{sec:systematics}}

\begin{figure}
    \centering
    \includegraphics[width=0.5\textwidth]{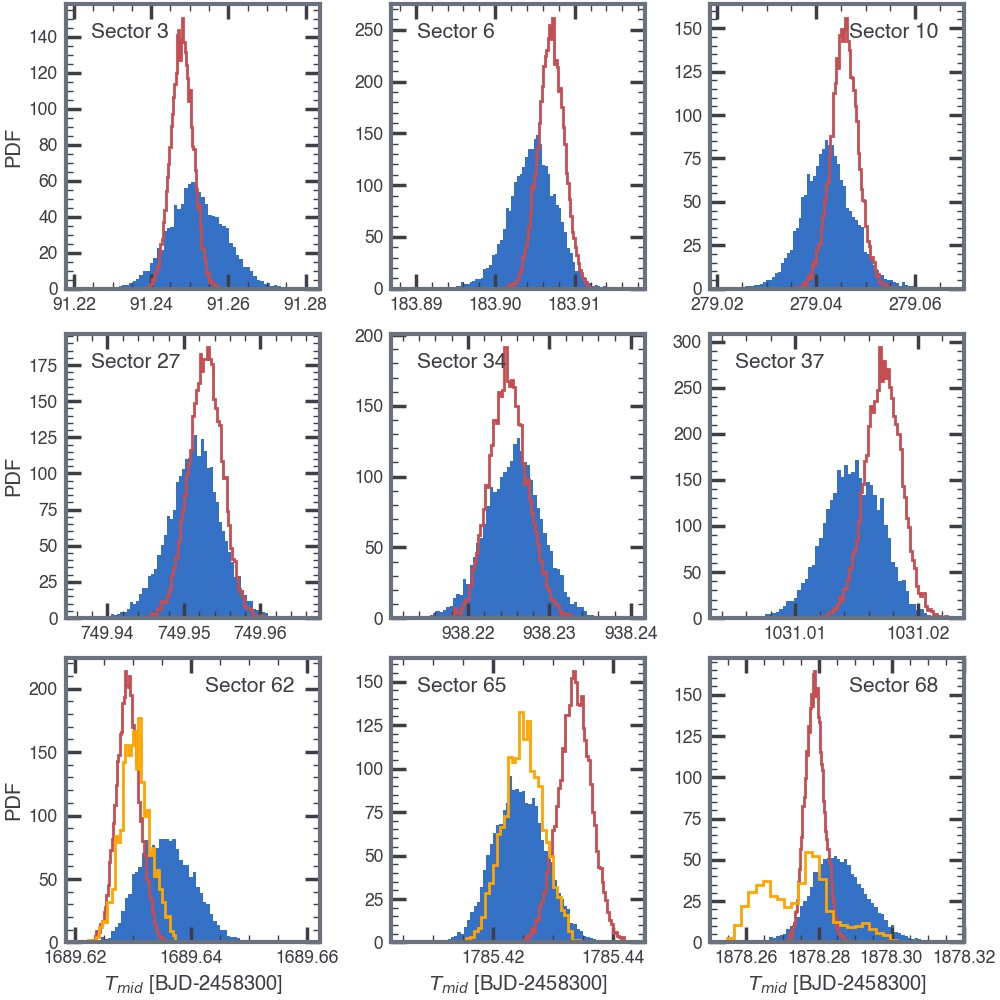}
    \caption{The comparisons between the fitted nine transit midtimes posteriors derived from direct fitting (blue, Section \ref{sec:transit_times}), photodynamical model (red, Section \ref{sec:sampling}), and Gaussian process correction (orange, see text in Section \ref{sec:systematics}). }
    \label{fig:midtime_posterior_compare}
\end{figure}

In Figure \ref{fig:midtime_posterior_compare}, we compare the posteriors of transit midtimes of all sectors from the direct fitting (Table \ref{tab:transit_time_duration}) and photodynamical fitting. Generally, the uncertainties of transit midtime from the global photodynamical model are smaller than those from direct fitting, and the midtime posteriors from two analyses are consistent within 1$\sigma$, except for the transits in S62, S65, and S68. 
The in-transit systematics that cannot be removed from the detrending technique described above could potentially bias the midtime estimates and thus the photodynamical modeling. 
To investigate whether the in-transit systematics, especially in S62 and S68, have a significant effect on the photodynamical modeling, we took an independent analysis of midtime fitting incorporating the Gaussian process (GP) to see if the systematic-corrected midtimes are consistent with the results from the photodynamical model. The GP could model the noise and transit altogether to yield more unbiased midtime results compared to the direct fitting. This is done by using the python package \texttt{juliet} \citep{Espinoza2019}. 

We show the GP-correct midtime posteriors in S62, S65 and S68 in Figure \ref{fig:midtime_posterior_compare}. For the S62 and S68, the midtimes derived from the photodynamical modeling are more discrepant with those from direct fitting but are more consistent with the results of GP-corrected midtimes. This indicates that our midtime posteriors from the global photodynamical modeling do not greatly suffer from the systematics in S62 and S68 as the direct fitting does. 
However, the transit in Sector 65 displays $\sim 2\sigma$ difference between photodynamical modeling and other results. We postulate that this is caused by the very sharp flux jump near the egress phase of this transit event (see Figure \ref{fig:best_fit}), which could not be sufficiently modeled by the GP. This could potentially make the predicted egress time from the direct fitting end earlier and thus the midtime will be a little earlier too.

\subsection{Stability Analysis \label{sec:stability}}

\begin{figure}
    \centering
    \includegraphics[width=0.45\textwidth]{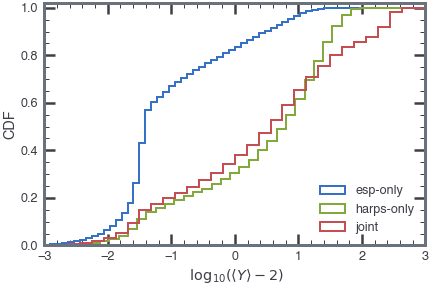}
    \caption{Dynamical stability (in terms of the displacement of MEGNO $\langle Y \rangle$ factor from 2) for three sets of photodynamical solutions from Table \ref{tab:model_bestfit}. Each solutions are integrated for 50 kyr. The closer the $\langle Y \rangle$ is to 2, the more stable the system is in the integrated time. The stability analysis shows that the majority of \textit{esp}-only solutions is stable while the other sets of solutions are more prone to be unstable within the integrated time, mainly owing to the different eccentricity of TOI-1338 c from the different solutions (see Section \ref{sec:stability}).}
    \label{fig:megno}
\end{figure}

To investigate the stability of the posterior solutions from our photodynamical modeling, we calculated a dynamical Mean Exponential Growth factor of Nearby Orbits \citep[MEGNO;][]{Gozdziewski2001,Cincotta2003} for all three sets of solutions in Table \ref{tab:model_bestfit}. 
    MEGNO is a numerical tool to differentiate chaotic and stable motion. Specifically, if the orbits are stable, the final $\langle Y \rangle$ will converge to 2, else it will increase as time. 
    We integrate each solution with the \texttt{whfast} \citep{Rein2015}
    integrator for 50 kyr and calculate the MEGNO value $\langle Y \rangle$ of the system. 
The cumulative density function of calculated $\langle Y \rangle$ for three sets of solutions are shown in Figure \ref{fig:megno}. 
    The median $\langle Y \rangle$ value of \textit{joint}-, \textit{esp}-, and \textit{harps}-only solutions are 6.13, 1.98, and 7.84, respectively. 
    This result suggests the majority of posterior solutions ($\sim 80\%$ for $1 < \langle Y \rangle < 3$) from \textit{esp}-only analysis is stable within 50 kyr, while the other two sets of posteriors are more likely to become unstable within the integrated time. 
    Examining the unstable solutions with $\langle Y \rangle > 3$ shows that the larger the eccentricity of $e_c$ TOI-1338 c is, the more likely the system would get unstable. 
Note that the \textit{joint} and \textit{harps}-only solutions yielded $\sim 0.1$ for $e_c$ while \textit{esp}-only solutions have $\sim 0.047$. This is also consistent with the conclusion of \citetalias{Standing2023} that the eccentricities of both planets should be lower than 0.1 to keep the system stable. 
Based on the results from the stability analysis, it is advised to take the \textit{esp}-only solutions as the final adopted solutions from our photodynamical modeling.

\section{Results \label{sec:photo_result}}

In our photo-dynamical modeling, we analyze the mass constraint of TOI-1338 b, which could potentially have an extremely low density as $0.137 \pm 0.026$ g/cm$^{-3}$, in Section \ref{sec:inner_planet_mass}. We also discuss the mutual inclination of non-transiting planet TOI-1338 c relative to the binary plane in Section \ref{sec:mutual_inclination}. Then we present the future primary transits forecast in Section \ref{sec:primary_transit_forecast}.

\subsection{Eclipse Timing Variation Analysis \label{sec:etv_analysis}}

The mass of the circumbinary planet is traditionally constrained by the eclipse timing variations \citep[e.g., ][]{Doyle2011, Welsh2012}. The planet will gravitationally perturb the binary and cause the apsidal precession of binary orbits, which would appear as a divergence of the binary period derived from primary and secondary eclipses if the binary orbit is eccentric. 
Futhermore, if the planet is non-coplanar with the binary, the inclination of binary orbit would also go through precession and manifest as the eclipse depth variations, though this would work most effectively for grazing eclipses \citep[see, e.g., ][]{Socia2020,Goldberg2023}. 
The planet's mass could also be constrained by the short-term ``choppings'' in the Observation minus Computed (O-C) diagram, the time-scale of which is comparable to planetary orbit \citep{Rappaport2013}. 

\begin{figure*}
    \centering
    \includegraphics[width=\textwidth]{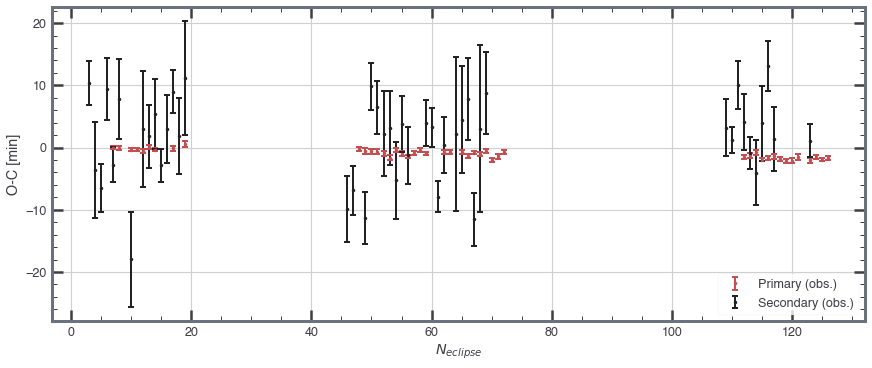}
    \caption{The observation minus the calculation (O-C) diagram of the midtimes of primary and secondary eclipses, fitted to the common period. The periods fitted from primary and secondary eclipses are 14.6085574$\pm$ 9$\times10^{-7}$ day and 14.6085774$\pm$ 1.58$\times 10^{-5}$ day, respectively. Neither the divergence of primary and secondary ephemerides (0.8 min) nor the dynamical delay of the primary eclipse due to the short-term perturbations (0.03 min) predicted from the best-fit model is observable given the large uncertainties in midtimes of primary and secondary eclipses (0.36 min and 5 min, respectively).}
    \label{fig:o-c_diagram}
\end{figure*}

In the best-fit model excluding GR, the total apsidal precession of binary orbit is 31.35 arcsec per year. The precession rate caused by planet b is 14.56 arcsec per year, slightly larger than the expected GR-induced precession rate ($\sim 9.01$ arcsec per year) and is around 10 times higher than the tidal-induced precession rate, but is smaller than the precession rate induced by the outer planet ($\sim 17.91$ arcsec per year). 
In the 5-year TESS's observation baseline, this precession rate will lead to a divergence of 0.8 minutes between primary and secondary eclipses when they are fitted to common ephemerides. However, this divergence amplitude is not readily observable because it is much smaller than the typical uncertainty of the secondary eclipse midtimes ($\sim$5-6 min), as seen from Figure \ref{fig:o-c_diagram}. 
The short time-scale oscillations of O-C curves have an amplitude of 0.03 min, dominated by the outer-planet perturbation. 
We also tried out a few cases when TOI-1338 c is misaligned by mutual inclination up to 40$^\circ$, and the amplitude of primary ETVs is still smaller than the typical uncertainties of 0.3 min. 
Therefore the ETVs are not likely to constrain the mass of TOI-1338 b nor the inclination of TOI-1338 c.

\subsection{The Mass of TOI-1338 b \label{sec:inner_planet_mass}}

Our adopted photodynamical solutions in Table \ref{tab:model_bestfit} show that the mass of TOI-1338 b is $11.3\pm2.1$ $M_\oplus$. To determine the source of the inner planetary mass constraint, we run two additional groups of modeling with modified fitted data to see whether the mass of TOI-1338 b is constrained or not. 
	The first group of modeling fits the stellar eclipse photometry in the first year of TESS observation (S3-S10) and the full set of RV observations, this choice aligns with \citetalias{Kostov2020} and removes the possible mass constraints delivered by long-term ETVs. However, the mass constraints for planet b remain the same as the adopted solutions. 
	The second group of modeling fits all TESS photometry but without RV data, with the mass of the binary and outer planet fixed at the best-fit value of the adopted solution. In this case, the mass distribution of the TOI-1338 b is highly skewed to zero, with a 95\% upper limit of $26.1~M_\oplus$. 
Therefore we conclude that the mass reported in Table \ref{tab:model_bestfit} are constrained from the RV data. 

\begin{figure*}
    \centering
    \includegraphics[width=0.95\textwidth]{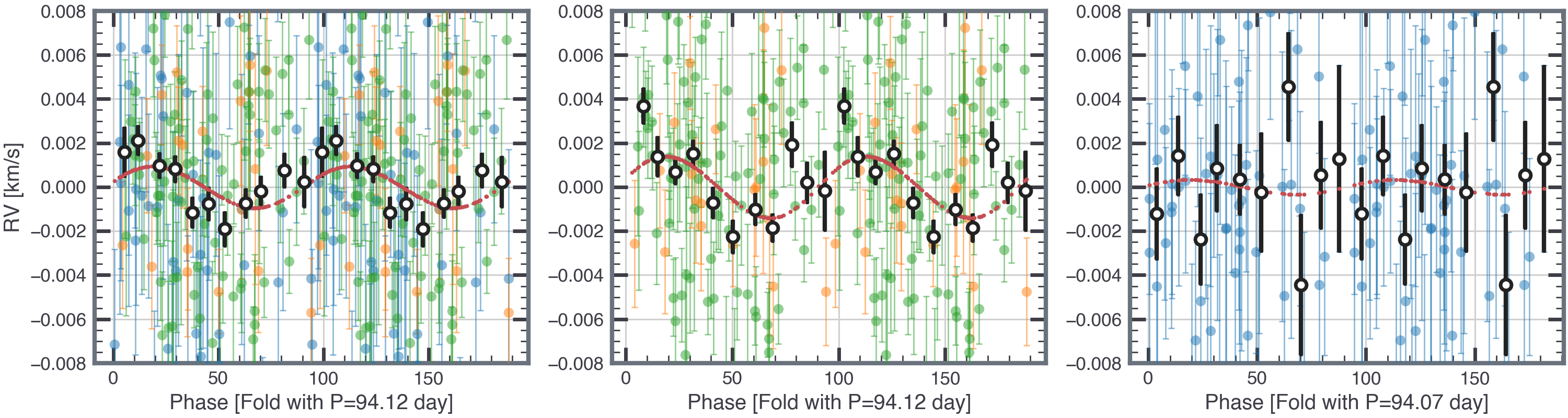}
    \caption{Left to right: the best-fit RV signals of TOI-1338 b from \textit{joint}, \textit{esp}-only, and \textit{harps}-only analyses. The RV data from HARPS, and ESPRESSO prior to and after 2019 are shown in blue, orange, and green, respectively. Two periods of TOI-1338 b are shown in the phase-folded figure. Black symbols with error bars  are RVs binned in 11 evenly-spaced intervals within one period. The simulated RV signals of TOI-1338 b in each analysis at the observation epoch are also plotted in red dots.} 
    \label{fig:rv_massb}
\end{figure*}

We then look for the RV signal of TOI-1338 b in the current RV dataset. 
This is done by subtracting the observed RVs from the secondary star and outer planet's RV signal simulated issued from the integrator, the parameters of which are taken from the best-fit solutions. The residuals should correspond to the RV signal of TOI-1338 b. 
The phase-folded residuals are shown in Figure \ref{fig:rv_massb}. 
    In the left and middle panel of Figure \ref{fig:rv_massb}, corresponding to the best-fit TOI-1338 b's RV signal from the \textit{joint} and \textit{esp}-only analyses, we observe clear modulation in observed RV residuals folded at the period of TOI-1338 b, with a semi-amplitude of $\sim 1.0$ and 1.5 m/s, respectively. 
    To assess the compatibility of this modulation with our model, we have included in Figure \ref{fig:rv_massb} the predicted RV signals of TOI-1338 b from the best-fit model at each of the RV observation epochs, represented as red points. The figure demonstrates a close match between the predicted signals and the observed modulations. 
It is important to note that some deviations of the observed RV signals from the model predictions are apparent, particularly in certain phase bins. These deviations may be attributed to the uneven coverage of RV observations over the planet phases, especially in \textit{esp}-only results where the current RV data are sparser in the latter half of the planetary period (phase $\sim 70-94$ day).


The \textit{harps}-only solution provides a mass constraint of TOI-1338 b consistent with zero. As is also seen in the right panel of Figure \ref{fig:rv_massb}, there is no apparent signal when the RVs from the secondary star and TOI-1338 c are subtracted. 
    After subtracting the best-fit signal of TOI-1338 b, the $\chi^2$ changed from 57 to 58, so there is no improvement made. 
It is likely that the RV signal of TOI-1338 b is completely submerged in the noise of HARPS RV measurements. Also note that in Section \ref{sec:sampling} we report inconsistency in binary RV signal between HARPS and ESPRESSO with amplitude $\sim 14$ m/s. Such large systematics could be detrimental to the search of planets with signals as weak as only $\sim$1 m/s.

The study of \citetalias{Standing2023} and \citetalias{Kostov2020} have both formerly provided mass constraints on TOI-1338 b. 
    \citetalias{Standing2023} uses RV datasets identical to ours and perform Keplerian fit to the RV and find tentative signals at periodicity around 100 days consistent with TOI-1338 b, though the significance of this signal is not high enough and thus they report a 99\% upper limit of $21.8 \pm 0.9~M_\oplus$ for TOI-1338 b. Note that their analysis is purely based on RV data, independent of any prior knowledge of orbital parameters TOI-1338 b, including the orbital phases which could be precisely informed by the transit events. 
    \citetalias{Kostov2020} derived $33\pm20~M_\oplus$ for the planet by photodynamical modeling of 1.5 yr TESS photometry and CORALIE/HARPS RV measurements. The mass constraint mainly comes from the seven HARPS RV measurements obtained at that time. 
Our method is more similar to the analysis of \citetalias{Kostov2020}, except that we incorporate more RV measurements from ESPRESSO with higher precision and yield more precise mass constraints.


\subsection{The Inclination of TOI-1338 c \label{sec:mutual_inclination}}

\begin{figure}
    \centering
    \includegraphics[width=0.45\textwidth]{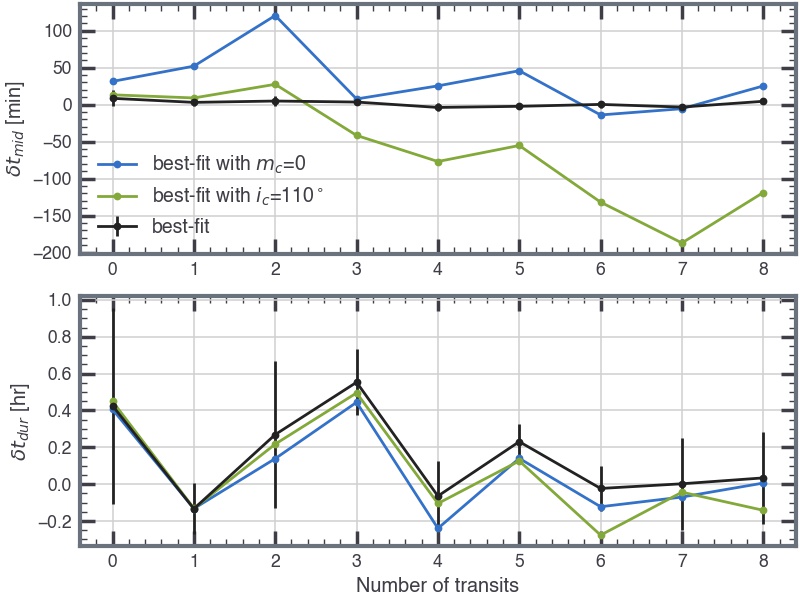}
    \caption{The difference between the measured transit midtimes and durations of TOI-1338 and those derived from best-fit solution (black points). The error bars are the measured uncertainties from the observation (listed in Table \ref{tab:transit_time_duration}). We also show how the transit midtimes/durations of TOI-1338 b would change if we tweaked the TOI-1338 c's planetary mass to be zero (blue) and inclination to be 110$^\circ$ (green), the deviations shown by the tweaked models indicate TOI-1338 b is gravitationally interacting with TOI-1338 c. }
    \label{fig:tmid_dur_different}
\end{figure}

\begin{figure}
    \centering
    \includegraphics[width=0.4\textwidth]{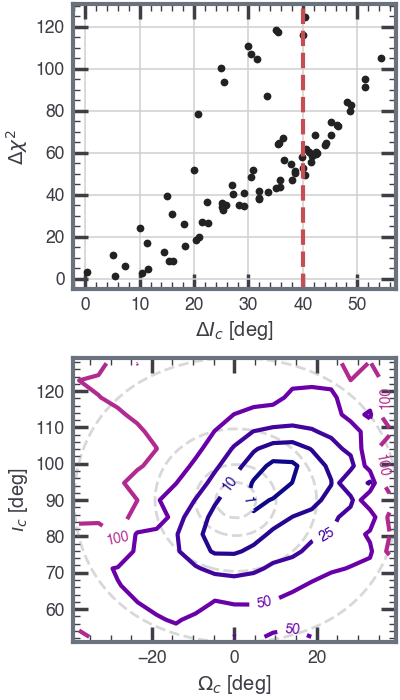}
    \caption{Top: optimized $\Delta \chi^2$ map of transit light curve of TOI-1338 b and RV data fitting as a function of the mutual inclination between TOI-1338 c and binary orbit $\Delta I_c$. The red vertical dashed line denotes the stability limit ($\sim 40^\circ$) derived from \citetalias{Standing2023}, beyond which the system would be unstable. Bottom: The isolines of optimized $\Delta \chi^2$ map of transit light curve and RV data fitting assuming different orbital inclination $i_c$ and node angles $\Omega_c$ of TOI-1338 c. The isolines of mutual inclinations of 5$^\circ$, 10$^\circ$, 20$^\circ$, and 40$^\circ$ are also plotted in gray dashed lines. }
    \label{fig:mutual_inclination_map}
\end{figure}

In Figure \ref{fig:tmid_dur_different}, we show evidence that the TOI-1338 b is gravitationally perturbed by TOI-1338 c, which manifests as TTVs and TDVs in models where we fixed the mass of TOI-1338 c as zero. 
The variations in transit timing are more prominent, in several transit events the deviations can be detected with several $\sigma$, whereas the variations in transit durations are close to the measured uncertainties. 
Moreover, the inclination of the outer planet's orbit also changes the TTVs, which may allow us to put meaningful three-dimensional constraints on the outer planet's orbit, which is inaccessible to the RV method by which TOI-1338 c is discovered.

The mutual inclination between the binary and outer TOI-1338 c can be calculated via
\begin{equation}
    \Delta I_c = \cos i_B \cos i_c + \sin i_B \sin i_c \cos\Omega_c
\end{equation}

The three analyses yielded a mutual inclination of TOI-1338 c around 10$^\circ$ relative to the binary orbital plane, consistent with a coplanar configuration. However, our adopted model prefers a non-transit nature of TOI-1338 c, even though we did not impose any criterion on it. 


To explore how sensitively the data constrain the mutual inclination of TOI-1338 c $\Delta I_c$, we run a suite of additional fits, with $i_c$ being between 50$^\circ$ and 130$^\circ$ and $\Omega_c$ ranging between -40$^\circ$ and 40$^\circ$ and a step of 5$^\circ$. 
 All other parameters are fixed at the best-fit value of \textit{joint} analyses in Table \ref{tab:model_bestfit}, except that $P_{b}$, $\sqrt{e_b}\cos\omega_b$, $\sqrt{e_b}\sin\omega_b$, $\lambda_b$, $\sqrt{e_c}\cos\omega_c$, $\sqrt{e_c}\sin\omega_c$, $\lambda_c$, and $P_c$ are further optimized by DE-MCMC.
 The inclination and node angle of the inner planet are not sensitive to the transit timing and are therefore kept fixed. These parameters are optimized against the observational dataset identical to the adopted solutions, except that  binary eclipse light curves are excluded since the ETVs are not sensitive to the inclination of TOI-1338 c (See Section \ref{sec:etv_analysis})
 The model optimization is carried out by DE-MCMC, with 40 chains for each experiment, and run for 4000-21000 steps (usually when the outer orbit is kept at high $\Delta I_c$, the optimization will need to run longer chains).
 Optimization ends when we visually confirm that the $\chi^2$ remains constant for more than 50\% of the steps.

The optimized $\chi^2$ map of the dynamical fitting is shown at the bottom of Figure \ref{fig:mutual_inclination_map}. We find the minimum $\chi^2$ exists in ($i_c$, $\Omega_c$)=(95$^\circ$, 10$^\circ$), consistent with the solutions provided in Table \ref{tab:model_bestfit}. That is, a lower mutual inclination of TOI-1338 c is favored by the current data, and the fits become worse in larger mutual inclinations, as shown in the top panel of Figure \ref{fig:mutual_inclination_map}. 
When examining the $\chi^2$ of transit light curves and RV data separately, we found that when TOI-1338 c is fixed at high $\Delta I_c$ the optimized results could provide a good match to the observed transit midtimes and durations, while the fits of RV data become worse. 

However, compared to the nominal photodynamical models in Section \ref{sec:photodynam},
 only the eclipse photometry is removed in the models in this section, therefore the degree of freedom of these model optimizations is still as high as 10336. At higher $\Delta I_c$, the $\Delta \chi^2$ is only $\sim$35-40, which may not be a statistically significant argument against high $\Delta I_c$ if seen from the perspective of reduced-$\chi^2$. 
Therefore, based on the current data, lower mutual inclination between binary and outer planet's orbit is marginally favored over high mutual inclination scenarios.

\subsection{Future Transits Forecast \label{sec:primary_transit_forecast}}

\begin{figure*}[ht]
    \centering
    \includegraphics[width=0.95\textwidth]{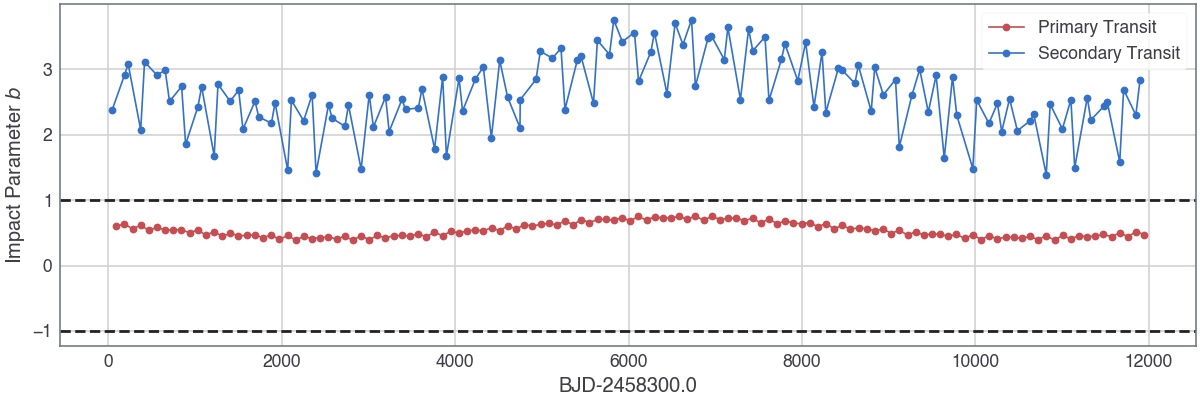}
    \includegraphics[width=\textwidth]{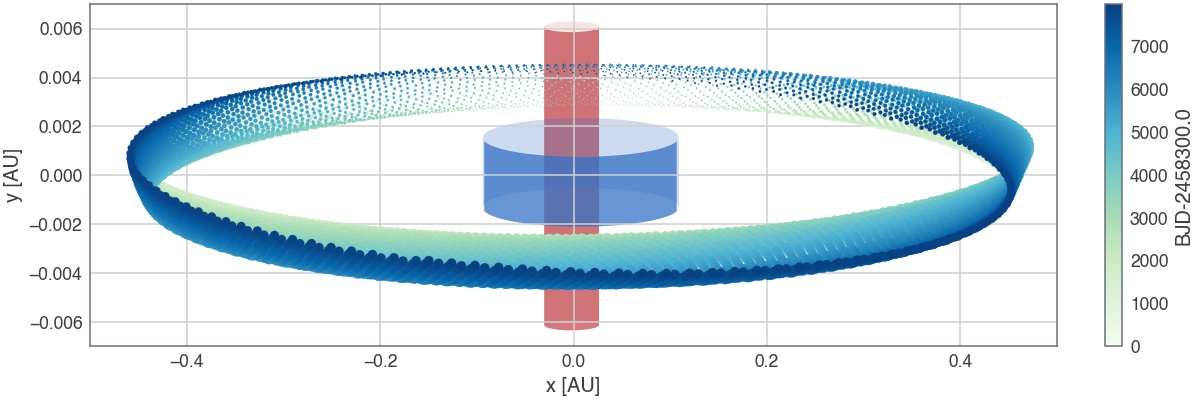}
    \caption{The evolution of orbital plane and the change of impact parameter of TOI-1338 b's primary and secondary transits. Top: the evolution of impact parameters of primary (red) and secondary transit (blue) integrated from the best-fit model. It shows that TOI-1338 b will permanently transit the primary star but hardly transit the secondary star. Bottom: the sky-projection view of the orbits of binary and TOI-1338 b. The orbit of TOI-1338 b is integrated 7000 days to show the effect of nodal precession due to interactions with binary stars. The size of the point scales with the line-of-sight distance, thus smaller points are behind the binary orbits. The red and blue regions show the crossing area of stellar disks of primary and secondary stars, accounting for the binary orbital motions. Transits can occur when the planet orbit and the stellar crossing area are intersected. The aspect ratio is 70:1. Due to the small mutual inclination of TOI-1338 b, the variation range of planetary inclination is very small, thus its orbit will always intersect the primary star's orbital crossing region. The secondary star has a smaller radius, thus the stellar and planetary orbits are not mutually intersected, precluding potential transitability.}
    \label{fig:impact_parameter}
\end{figure*}

We present the future primary transit of TOI-1338 b in Table \ref{tab:primary_transit_forecast}. It shows that TOI-1338 b will permanently transit the primary star, which is similar to Kepler-47 b \citep{Orosz2019}. This is partly due to the coplanarity of the orbit of TOI-1338 b relative to the binary planet ($\Delta I_b \sim 0.12^\circ$) and the sufficiently large primary radius. 
Integrating the best-fit model with \texttt{rebound}, in Figure \ref{fig:impact_parameter} we present the impact parameter of TOI-1338 b in each conjunction with the primary and the secondary star. Due to perturbation of the inner binary, the orbital plane of TOI-1338 b will undergo precession and the inclination of TOI-1338 b will follow sinusoidal-like variations.

Our best-fit models suggest that no secondary transit occurs in the current TESS observation. Following \cite{Martin2015a}, the minimum mutual inclination of TOI-1338 b ever transiting secondary star is 0.277$^\circ$, whereas our posterior samples yielded a $0.127^{+0.025 \circ}_{-0.024}$. This partly explains why secondary transits of TOI-1338 b are rare in our posterior models. The orbit of TOI-1338 b will hardly precess into an orientation that the planetary and secondary star's orbits are mutually intersected.

With a TESS mag of 11.45 mag and a period of 95 days, TOI-1338 b is an ideal circumbinary planet target for JWST transmission observation.  TESS is scheduled to re-observed TOI-1338 in Year 7 observation, and another three primary transits of TOI-1338 b are expected to occur in the observing window of Sector 89, 93, and 96 between 2025 Match and September. We encourage more follow-up observations to be carried out to TOI-1338 b transits to update and refine future transit ephemerides.

\section{Dicussion and Conclusion}
\label{sec:conclusion}

\begin{figure}[ht]
    \hspace{-0.55cm}
    \centering
    \includegraphics[width=0.50\textwidth]{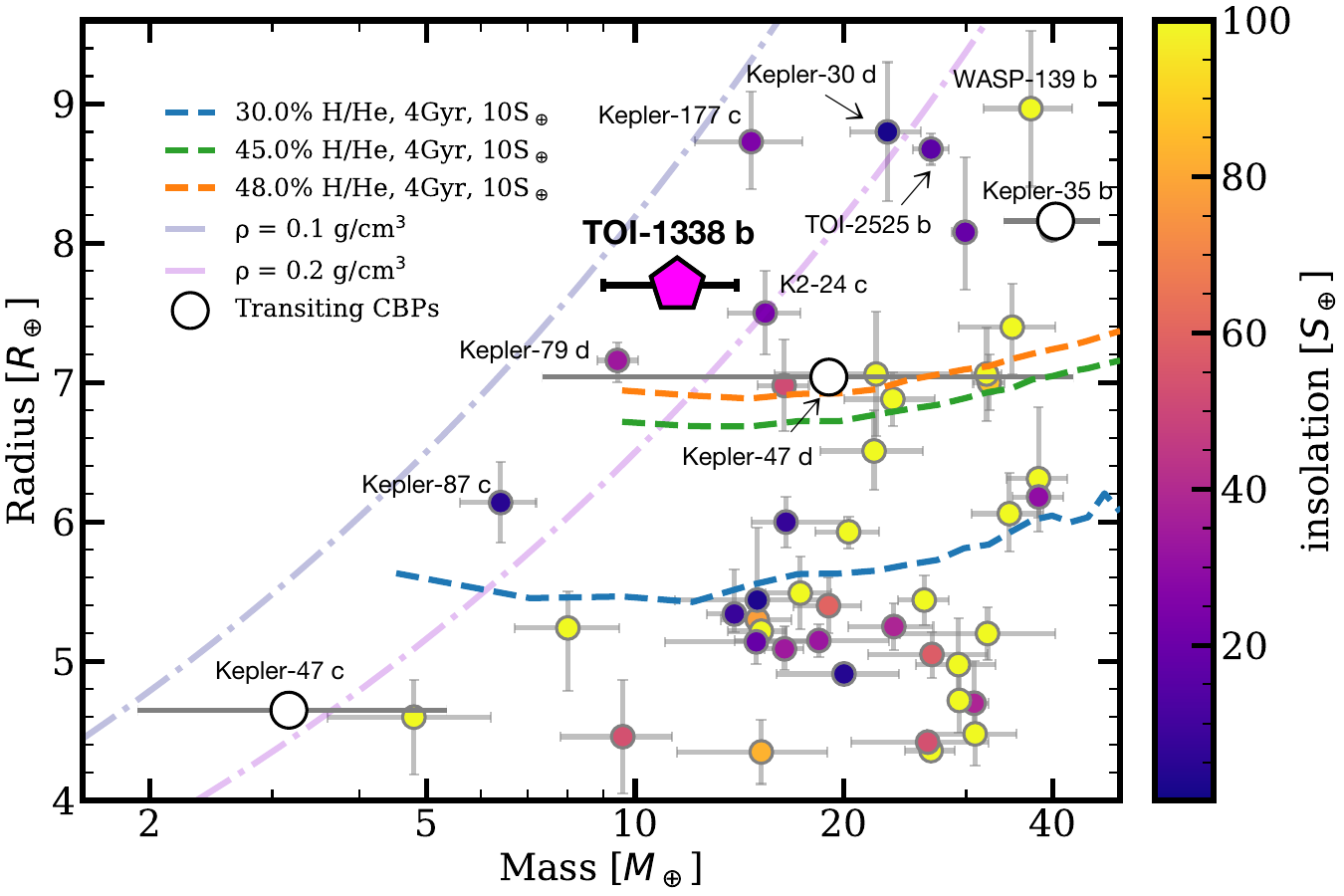}
    \caption{The masses and radii of low-mass sub-Jovian planets ($4R_\oplus<R_p< 9R_\oplus, 2 M_\oplus<M_p<40 M_\oplus$) on the NASA Exoplanet Archive as of 2024 April 7 with masses and radius measured better than 30\%, color-coded by the insolation fluxes. TOI-1338 b is highlighted in magenta. We also plot the transiting circumbinary planets in the sub-Saturn regime with larger white-filled circles. Dashed lines show the expected mass and radius for sub-Saturn planets at 4 Gyr old and 10 Earth insolation S$_\oplus$ from the tide-free, thermal evolution models of \cite{Millholland2020}, for H/He envelope mass fraction of 30\%(blue), 45\%(green), and 48\%(orange).}
    \label{fig:toi_1338_mr_diagram}
\end{figure}

In this work, we perform a system update of the TOI-1338 with two circumbinary planets. Combined with the latest TESS light curves and published radial velocity data, we carry out photodynamical analysis to self-consistently account for the mutual interactions between bodies within the systems. 

We determined a coplanar configuration for TOI-1338 c relative to the binary plane $\Delta I_c=9.1^{+6.0 \circ}_{-4.8}$, with a 99\% upper limit of $22^\circ$. Therefore the genuine mass of TOI-1338 c is determined to be $75.4^{+4.0}_{-3.6}$ M$_\oplus$.
This constraint mainly comes from the additional number of observed transits of TOI-1338 b and the precise RV measurements, where the gravitational pull from the outer planet starts to manifest in the transit timing variations of the inner planet (see Figure \ref{fig:mutual_inclination_map}). 
Together with the spin-orbit alignment between the primary star and the orbit of TOI-1338 b (projected spin-orbit angle of $2. ^\circ 8 \pm 17.^\circ 1$) \citep{Hodzic2020}, the whole system is in an alignment between primary's spin axis and binary-planet orbit configuration. This suggests the whole system forms from a single disk with no breaking or warps. 

 With our refined mass ($11.3 \pm 2.1$ M$_\oplus$) and radius ($7.66 \pm 0.053$ R$_\oplus$) for TOI-1338 b, we derived a low bulk density of $0.137 \pm 0.026$ g/cm$^{-3}$. This is comparable to circumbinary planet Kepler-47 c which also has a low bulk density of 0.17 g/cm$^{-3}$. In Figure \ref{fig:toi_1338_mr_diagram} we show TOI-1338 b and other low-mass sub-Saturns in the mass-radius diagram, with the tide-free thermal evolution models for sub-Saturns from \cite{Millholland2020} at TOI-1338 b's age ($\sim$4 Gyr, \citetalias{Kostov2020}) and insolation. We estimate the envelope mass fraction, f$_{\rm env}$, to exceed 50\%, implying that TOI-1338 b possesses a substantial envelope, akin to several super-puffs found around single stars (e.g., Kepler-79 d with f$_{\rm env}=49.1 \pm 1.9\%$, \cite{Millholland2020}; K2-24 c with f$_{\rm env}=52^{+5}_{-3}\%$, \cite{Petigura2018}).
\cite{Lee2016} suggests planets with such low density might have migrated from a distant region where the disk is cool and has low opacity, enabling efficient cooling of the accretion and forming a massive envelope even with a low-mass core. This hypothesis aligns with the general picture that most observed CBPs may have undergone migration during their formation process \citep[e.g., ][]{Pierens2013,Coleman2023a,Coleman2023b}, due to the local low growth rate of planetary cores near the stability limit, which hinders the \textit{in situ} formation of CBPs \citep{Pierens2020,Pierens2021}.

With a TESS-band magnitude of 11.45 and a J-band magnitude of 10.95, TOI-1338 is an optimal target for future follow-up and transmission observation. Taking an average equilibrium temperature of 495 K for TOI-1338 b, we calculated the transmission spectroscopy metric of TOI-1338 b to be 86$\pm$17, which is slightly below the recommended threshold of 96 for sub-Jovian planets \citep{Kempton2018}. Although previous measurements show featureless transmission spectra for low-density sub-Saturns \citep{Libby-Roberts2020,Chachan2020}, hypothetically owing to the high altitude hazes or circumstellar rings \citep{Wang2019,Gao2020,Ohno2022}, it would still be crucial to test this scenario in the infrared band to see if the hazes or rings are genuinely present and inflate the apparent radius of the planet.

We provided an updated future primary transit ephemerides of TOI-1338 b in Table \ref{tab:primary_transit_forecast}. In our posterior models, TOI-1338 b will permanently transit the primary star at all nodal precession phases, owing to the extreme coplanarity with the binary plane and the large radius of the primary star.  We encourage more follow-up observations of TOI-1338 b's primary transit to refine the system parameters.

\begin{deluxetable*}{CrrCC}
\tablecaption{ Primary Transit Forecast of TOI-1338 b }
\label{tab:primary_transit_forecast}
\tablehead{
\colhead{Transit Midtime} &
\colhead{Year} & 
\colhead{UTC} &
\colhead{Duration } &
\colhead{Impact Parameter} \\
\colhead{(BJD—2,455,000)} &
\colhead{} & 
\colhead{} &
\colhead{(hour)} &
\colhead{}
}
\decimals
\startdata
 $5367.0935 \pm 0.0089$ & 2024-02-26 & 14:14:38.8 & $6.07 \pm 0.07$ & $0.450 \pm 0.027$ \\ 
 $5460.8325 \pm 0.0182$ & 2024-05-30 & 07:58:48.0 & $13.31 \pm 0.11$ & $0.386 \pm 0.023$ \\ 
 $5555.8630 \pm 0.0060$ & 2024-09-02 & 08:42:39.7 & $6.60 \pm 0.10$ & $0.435 \pm 0.035$ \\ 
 $5648.6132 \pm 0.0113$ & 2024-12-04 & 02:43:02.5 & $10.06 \pm 0.12$ & $0.402 \pm 0.030$ \\ 
 $5744.4534 \pm 0.0104$ & 2025-03-09 & 22:52:56.4 & $7.84 \pm 0.14$ & $0.411 \pm 0.042$ \\ 
 $5836.9195 \pm 0.0132$ & 2025-06-10 & 10:04:01.9 & $7.39 \pm 0.13$ & $0.420 \pm 0.040$ \\ 
 $5932.7322 \pm 0.0183$ & 2025-09-14 & 05:34:19.7 & $10.00 \pm 0.19$ & $0.389 \pm 0.048$ \\ 
 $6025.5927 \pm 0.0159$ & 2025-12-16 & 02:13:28.3 & $6.31 \pm 0.15$ & $0.428 \pm 0.050$ \\ 
 $6120.6154 \pm 0.0333$ & 2026-03-21 & 02:46:09.6 & $12.85 \pm 0.25$ & $0.370 \pm 0.053$ \\ 
 $6214.3975 \pm 0.0188$ & 2026-06-22 & 21:32:23.9 & $6.14 \pm 0.18$ & $0.431 \pm 0.061$ \\ 
 $6308.1259 \pm 0.0468$ & 2026-09-24 & 15:01:15.1 & $13.48 \pm 0.29$ & $0.371 \pm 0.058$ \\ 
 $6403.1374 \pm 0.0272$ & 2026-12-28 & 15:17:48.8 & $6.66 \pm 0.23$ & $0.430 \pm 0.071$ \\ 
 $6495.8436 \pm 0.0493$ & 2027-03-31 & 08:14:46.3 & $10.04 \pm 0.29$ & $0.396 \pm 0.066$ \\ 
 $6591.6908 \pm 0.0409$ & 2027-07-05 & 04:34:41.4 & $7.93 \pm 0.30$ & $0.420 \pm 0.079$ \\ 
 $6684.1435 \pm 0.0432$ & 2027-10-05 & 15:26:36.1 & $7.33 \pm 0.28$ & $0.429 \pm 0.078$ \\ 
 $6779.9414 \pm 0.0614$ & 2028-01-09 & 10:35:34.8 & $10.15 \pm 0.42$ & $0.412 \pm 0.086$ \\ 
 $6872.8293 \pm 0.0453$ & 2028-04-11 & 07:54:09.5 & $6.26 \pm 0.29$ & $0.451 \pm 0.088$ \\ 
 $6967.7647 \pm 0.0998$ & 2028-07-15 & 06:21:08.1 & $13.06 \pm 0.54$ & $0.402 \pm 0.090$ \\ 
 $7061.5889 \pm 0.0534$ & 2028-10-17 & 02:08:04.3 & $6.10 \pm 0.34$ & $0.470 \pm 0.099$ \\ 
 $7155.1419 \pm 0.1270$ & 2029-01-18 & 15:24:22.0 & $13.22 \pm 0.59$ & $0.414 \pm 0.095$ \\ 
 $7250.2715 \pm 0.0711$ & 2029-04-23 & 18:30:58.1 & $6.67 \pm 0.42$ & $0.480 \pm 0.106$ \\ 
 $7342.8505 \pm 0.1095$ & 2029-07-25 & 08:24:41.9 & $9.39 \pm 0.53$ & $0.452 \pm 0.103$ \\ 
 $7438.7453 \pm 0.0969$ & 2029-10-29 & 05:53:12.8 & $8.07 \pm 0.54$ & $0.482 \pm 0.110$ \\ 
 $7531.2057 \pm 0.0921$ & 2030-01-29 & 16:56:13.1 & $6.91 \pm 0.48$ & $0.494 \pm 0.112$ \\ 
 $7626.8889 \pm 0.1451$ & 2030-05-05 & 09:20:03.2 & $10.49 \pm 0.74$ & $0.480 \pm 0.113$ \\ 
 $7719.8762 \pm 0.0930$ & 2030-08-06 & 09:01:46.7 & $5.98 \pm 0.48$ & $0.527 \pm 0.119$ \\ 
 $7814.4865 \pm 0.2193$ & 2030-11-08 & 23:40:30.5 & $13.36 \pm 0.92$ & $0.478 \pm 0.114$ \\
\enddata
\end{deluxetable*}

\section*{Acknowledgments}

We thank the referee, Hans J. Deeg, whose comments greatly improved the quality and clarity of the manuscript. We also thank Fei Dai, Xian-Yu Wang, Sharon X. Wang, and Yuan-Zhe Dai for their helpful discussions. This work is supported by the National Natural Science Foundation of China (grant Nos. 11973028, 11933001, 1803012, 12150009) and the National Key R\&D Program of China (2019YFA0706601). We also acknowledge the science research grants from the China Manned Space Project with No. CMSCSST-2021-B12 and CMS-CSST-2021-B09, as well as the Civil Aerospace Technology Research Project (D050105). 

This paper includes data collected with the TESS mission, obtained from the MAST data archive at the Space Telescope Science Institute (STScI). Funding for the TESS mission is provided by the NASA Explorer Program. STScI is operated by the Association of Universities for Research in Astronomy, Inc., under NASA contract NAS 5-26555.  The TESS data presented in this paper were obtained from the Mikulski Archive for Space Telescopes (MAST) at the Space Telescope Science Institute. The TESS light curves used in this work can be accessed via MAST \citep{https://doi.org/10.17909/t9-nmc8-f686}. The GSFC TESS light curves can be accessed in \cite{https://doi.org/10.17909/j2yt-t417}.

\vspace{5mm}

\bibliographystyle{aasjournal}
\bibliography{ms}

\begin{thebibliography}{}
\expandafter\ifx\csname natexlab\endcsname\relax\def\natexlab#1{#1}\fi
\providecommand{\url}[1]{\href{#1}{#1}}
\providecommand{\dodoi}[1]{doi:~\href{http://doi.org/#1}{\nolinkurl{#1}}}
\providecommand{\doeprint}[1]{\href{http://ascl.net/#1}{\nolinkurl{http://ascl.net/#1}}}
\providecommand{\doarXiv}[1]{\href{https://arxiv.org/abs/#1}{\nolinkurl{https://arxiv.org/abs/#1}}}

\bibitem[{{Aller} {et~al.}(2020){Aller}, {Lillo-Box}, {Jones}, {Miranda}, \&
  {Barcel{\'o} Forteza}}]{Aller2020}
{Aller}, A., {Lillo-Box}, J., {Jones}, D., {Miranda}, L.~F., \& {Barcel{\'o}
  Forteza}, S. 2020, \aap, 635, A128, \dodoi{10.1051/0004-6361/201937118}

\bibitem[{Almenara {et~al.}(2022)Almenara, H{\'{e}}brard, D{\'{i}}az, Laskar,
  Correia, Anderson, Boisse, Bonfils, Brown, Casanova, {Collier Cameron},
  Fern{\'{a}}ndez, Jenkins, Kiefer, {Lecavelier Des Etangs}, Lissauer,
  MacIejewski, McCormac, Osborn, Pollacco, Ricker, S{\'{a}}nchez, Seager, Udry,
  Verilhac, \& Winn}]{Almenara2022}
Almenara, J.~M., H{\'{e}}brard, G., D{\'{i}}az, R.~F., {et~al.} 2022, Astronomy
  and Astrophysics, 663, \dodoi{10.1051/0004-6361/202142964}

\bibitem[{Armstrong {et~al.}(2013)Armstrong, Martin, Brown, Faedi, Chew,
  Mardling, Pollacco, Triaud, \& Udry}]{Armstrong2013}
Armstrong, D., Martin, D.~V., Brown, G., {et~al.} 2013, Monthly Notices of the
  Royal Astronomical Society, 434, 3047, \dodoi{10.1093/mnras/stt1226}

\bibitem[{Armstrong {et~al.}(2014)Armstrong, Osborn, Brown, Faedi, {G{\'{o}}mez
  Maqueo Chew}, Martin, Pollacco, \& Udry}]{Armstrong2014}
Armstrong, D.~J., Osborn, H.~P., Brown, D.~J., {et~al.} 2014, Monthly Notices
  of the Royal Astronomical Society, 444, 1873, \dodoi{10.1093/mnras/stu1570}

\bibitem[{Brooks \& Gelman(1998)}]{Brooks1998}
Brooks, S.~P., \& Gelman, A. 1998, Journal of Computational and Graphical
  Statistics, 7, 434, \dodoi{10.1080/10618600.1998.10474787}

\bibitem[{Chachan {et~al.}(2020)Chachan, Jontof-Hutter, Knutson, Adams, Gao,
  Benneke, Berta-Thompson, Dai, Deming, Ford, Lee, Libby-Roberts, Madhusudhan,
  Wakeford, \& Wong}]{Chachan2020}
Chachan, Y., Jontof-Hutter, D., Knutson, H.~A., {et~al.} 2020, The Astronomical
  Journal, 160, 201, \dodoi{10.3847/1538-3881/abb23a}

\bibitem[{Chen \& Kipping(2021)}]{Chen2021}
Chen, Z., \& Kipping, D. 2021, 12, 1.
\newblock \doarXiv{2112.00966}

\bibitem[{Cincotta {et~al.}(2003)Cincotta, Giordano, \&
  Sim{\'{o}}}]{Cincotta2003}
Cincotta, P.~M., Giordano, C.~M., \& Sim{\'{o}}, C. 2003, Physica D: Nonlinear
  Phenomena, 182, 151, \dodoi{10.1016/S0167-2789(03)00103-9}

\bibitem[{{Claret}(2017)}]{Claret2017}
{Claret}, A. 2017, \aap, 600, A30, \dodoi{10.1051/0004-6361/201629705}

\bibitem[{Coleman {et~al.}(2023{\natexlab{a}})Coleman, Nelson, \&
  Triaud}]{Coleman2023a}
Coleman, G.~A., Nelson, R.~P., \& Triaud, A.~H. 2023{\natexlab{a}}, Monthly
  Notices of the Royal Astronomical Society, 522, 4352,
  \dodoi{10.1093/mnras/stad833}

\bibitem[{Coleman {et~al.}(2023{\natexlab{b}})Coleman, Nelson, Triaud, \&
  Standing}]{Coleman2023b}
Coleman, G. A.~L., Nelson, R.~P., Triaud, A. H. M.~J., \& Standing, M.~R.
  2023{\natexlab{b}}, Monthly Notices of the Royal Astronomical Society, 527,
  414, \dodoi{10.1093/mnras/stad3216}

\bibitem[{{Correia} {et~al.}(2013){Correia}, {Bou{\'e}}, {Laskar}, \&
  {Morais}}]{correia2013}
{Correia}, A.~C.~M., {Bou{\'e}}, G., {Laskar}, J., \& {Morais}, M.~H.~M. 2013,
  \aap, 553, A39, \dodoi{10.1051/0004-6361/201220482}

\bibitem[{Dawson {et~al.}(2014)Dawson, Johnson, Fabrycky, Foreman-Mackey,
  Murray-Clay, Buchhave, Cargile, Clubb, Fulton, Hebb, Howard, Huber, Shporer,
  \& Valenti}]{Dawson2014}
Dawson, R.~I., Johnson, J.~A., Fabrycky, D.~C., {et~al.} 2014, Astrophysical
  Journal, 791, \dodoi{10.1088/0004-637X/791/2/89}

\bibitem[{Doyle {et~al.}(2011{\natexlab{a}})Doyle, Carter, Fabrycky, Slawson,
  Howell, Winn, Orosz, Přsa, Welsh, Quinn, Latham, Torres, Buchhave, Marcy,
  Fortney, Shporer, Ford, Lissauer, Ragozzine, Rucker, Batalha, Jenkins,
  Borucki, Koch, Middour, Hall, McCauliff, Fanelli, Quintana, Holman, Caldwell,
  Still, Stefanik, Brown, Esquerdo, Tang, Furesz, Geary, Berlind, Calkins,
  Short, Steffen, Sasselov, Dunham, Cochran, Boss, Haas, Buzasi, \&
  Fischer}]{Doyle2011}
Doyle, L.~R., Carter, J.~A., Fabrycky, D.~C., {et~al.} 2011{\natexlab{a}},
  Science, 333, 1602, \dodoi{10.1126/science.1210923}

\bibitem[{Doyle {et~al.}(2011{\natexlab{b}})Doyle, Carter, Fabrycky, Slawson,
  Howell, Winn, Orosz, Přsa, Welsh, Quinn, Latham, Torres, Buchhave, Marcy,
  Fortney, Shporer, Ford, Lissauer, Ragozzine, Rucker, Batalha, Jenkins,
  Borucki, Koch, Middour, Hall, McCauliff, Fanelli, Quintana, Holman, Caldwell,
  Still, Stefanik, Brown, Esquerdo, Tang, Furesz, Geary, Berlind, Calkins,
  Short, Steffen, Sasselov, Dunham, Cochran, Boss, Haas, Buzasi, \&
  Fischer}]{Doyle2011a}
---. 2011{\natexlab{b}}, Science, 333, 1602, \dodoi{10.1126/science.1210923}

\bibitem[{{Er} {et~al.}(2021){Er}, {{\"O}zd{\"o}nmez}, \& {Nasiroglu}}]{Er2021}
{Er}, H., {{\"O}zd{\"o}nmez}, A., \& {Nasiroglu}, I. 2021, \mnras, 507, 809,
  \dodoi{10.1093/mnras/stab2054}

\bibitem[{{Esmer} {et~al.}(2022){Esmer}, {Ba{\c{s}}t{\"u}rk}, {Selam}, \&
  {Ali{\c{s}}}}]{Esmer2022}
{Esmer}, E.~M., {Ba{\c{s}}t{\"u}rk}, {\"O}., {Selam}, S.~O., \& {Ali{\c{s}}},
  S. 2022, \mnras, 511, 5207, \dodoi{10.1093/mnras/stac357}

\bibitem[{{Espinoza} {et~al.}(2019){Espinoza}, {Kossakowski}, \&
  {Brahm}}]{Espinoza2019}
{Espinoza}, N., {Kossakowski}, D., \& {Brahm}, R. 2019, \mnras, 490, 2262,
  \dodoi{10.1093/mnras/stz2688}

\bibitem[{Gao \& Zhang(2020)}]{Gao2020}
Gao, P., \& Zhang, X. 2020, The Astrophysical Journal, 890, 93,
  \dodoi{10.3847/1538-4357/ab6a9b}

\bibitem[{Goldberg {et~al.}(2023)Goldberg, Fabrycky, Martin, Albrecht, Deeg, \&
  Nowak}]{Goldberg2023}
Goldberg, M., Fabrycky, D., Martin, D.~V., {et~al.} 2023, 12, 1.
\newblock \doarXiv{2308.09255}

\bibitem[{{Go{\'z}dziewski} {et~al.}(2001){Go{\'z}dziewski}, {Bois},
  {Maciejewski}, \& {Kiseleva-Eggleton}}]{Gozdziewski2001}
{Go{\'z}dziewski}, K., {Bois}, E., {Maciejewski}, A.~J., \&
  {Kiseleva-Eggleton}, L. 2001, \aap, 378, 569,
  \dodoi{10.1051/0004-6361:20011189}

\bibitem[{Hodzic {et~al.}(2020)Hodzic, Triaud, Martin, Fabrycky, Cegla,
  Cameron, Gill, Hellier, Kostov, Maxted, Orosz, Pepe, Pollacco, Queloz,
  S{\'{e}}gransan, Udry, \& Welsh}]{Hodzic2020}
Hodzic, V.~K., Triaud, A.~H., Martin, D.~V., {et~al.} 2020, Monthly Notices of
  the Royal Astronomical Society, 497, 1627, \dodoi{10.1093/mnras/staa2071}

\bibitem[{{Jenkins} {et~al.}(2016){Jenkins}, {Twicken}, {McCauliff},
  {Campbell}, {Sanderfer}, {Lung}, {Mansouri-Samani}, {Girouard}, {Tenenbaum},
  {Klaus}, {Smith}, {Caldwell}, {Chacon}, {Henze}, {Heiges}, {Latham},
  {Morgan}, {Swade}, {Rinehart}, \& {Vanderspek}}]{Jenkins2016}
{Jenkins}, J.~M., {Twicken}, J.~D., {McCauliff}, S., {et~al.} 2016, in Society
  of Photo-Optical Instrumentation Engineers (SPIE) Conference Series, Vol.
  9913, Software and Cyberinfrastructure for Astronomy IV, ed. G.~{Chiozzi} \&
  J.~C. {Guzman}, 99133E, \dodoi{10.1117/12.2233418}

\bibitem[{{Kempton} {et~al.}(2018){Kempton}, {Bean}, {Louie}, {Deming}, {Koll},
  {Mansfield}, {Christiansen}, {L{\'o}pez-Morales}, {Swain}, {Zellem},
  {Ballard}, {Barclay}, {Barstow}, {Batalha}, {Beatty}, {Berta-Thompson},
  {Birkby}, {Buchhave}, {Charbonneau}, {Cowan}, {Crossfield}, {de Val-Borro},
  {Doyon}, {Dragomir}, {Gaidos}, {Heng}, {Hu}, {Kane}, {Kreidberg}, {Mallonn},
  {Morley}, {Narita}, {Nascimbeni}, {Pall{\'e}}, {Quintana}, {Rauscher},
  {Seager}, {Shkolnik}, {Sing}, {Sozzetti}, {Stassun}, {Valenti}, \& {von
  Essen}}]{Kempton2018}
{Kempton}, E. M.~R., {Bean}, J.~L., {Louie}, D.~R., {et~al.} 2018, \pasp, 130,
  114401, \dodoi{10.1088/1538-3873/aadf6f}

\bibitem[{Kipping(2013)}]{Kipping2013}
Kipping, D.~M. 2013, Monthly Notices of the Royal Astronomical Society, 435,
  2152, \dodoi{10.1093/mnras/stt1435}

\bibitem[{Konacki {et~al.}(2010)Konacki, Muterspaugh, Kulkarni, \&
  He{\l}lminiak}]{Konacki2010}
Konacki, M., Muterspaugh, M.~W., Kulkarni, S.~R., \& He{\l}lminiak, K.~G. 2010,
  Astrophysical Journal, 719, 1293, \dodoi{10.1088/0004-637X/719/2/1293}

\bibitem[{Kostov {et~al.}(2013)Kostov, McCullough, Hinse, Tsvetanov,
  H{\'{e}}brard, D{\'{i}}az, Deleuil, \& Valenti}]{Kostov2013}
Kostov, V.~B., McCullough, P.~R., Hinse, T.~C., {et~al.} 2013, Astrophysical
  Journal, 770, \dodoi{10.1088/0004-637X/770/1/52}

\bibitem[{Kostov {et~al.}(2014)Kostov, McCullough, Carter, Deleuil, D{\'{i}}az,
  Fabrycky, H{\'{e}}brard, Hinse, Mazeh, Orosz, Tsvetanov, \&
  Welsh}]{Kostov2014}
Kostov, V.~B., McCullough, P.~R., Carter, J.~A., {et~al.} 2014, Astrophysical
  Journal, 784, \dodoi{10.1088/0004-637X/784/1/14}

\bibitem[{{Kostov} {et~al.}(2020){Kostov}, {Orosz}, {Feinstein}, {Welsh},
  {Cukier}, {Haghighipour}, {Quarles}, {Martin}, {Montet}, {Torres}, {Triaud},
  {Barclay}, {Boyd}, {Briceno}, {Cameron}, {Correia}, {Gilbert}, {Gill},
  {Gillon}, {Haqq-Misra}, {Hellier}, {Dressing}, {Fabrycky}, {Furesz},
  {Jenkins}, {Kane}, {Kopparapu}, {Hod{\v{z}}i{\'c}}, {Latham}, {Law},
  {Levine}, {Li}, {Lintott}, {Lissauer}, {Mann}, {Mazeh}, {Mardling}, {Maxted},
  {Eisner}, {Pepe}, {Pepper}, {Pollacco}, {Quinn}, {Quintana}, {Rowe},
  {Ricker}, {Rose}, {Seager}, {Santerne}, {S{\'e}gransan}, {Short}, {Smith},
  {Standing}, {Tokovinin}, {Trifonov}, {Turner}, {Twicken}, {Udry},
  {Vanderspek}, {Winn}, {Wolf}, {Ziegler}, {Ansorge}, {Barnet}, {Bergeron},
  {Huten}, {Pappa}, \& {van der Straeten}}]{Kostov2020}
{Kostov}, V.~B., {Orosz}, J.~A., {Feinstein}, A.~D., {et~al.} 2020, \aj, 159,
  253, \dodoi{10.3847/1538-3881/ab8a48}

\bibitem[{{Kostov} {et~al.}(2021){Kostov}, {Powell}, {Orosz}, {Welsh},
  {Cochran}, {Collins}, {Endl}, {Hellier}, {Latham}, {MacQueen}, {Pepper},
  {Quarles}, {Sairam}, {Torres}, {Wilson}, {Bergeron}, {Boyce}, {Bieryla},
  {Buchheim}, {Ben Christiansen}, {Ciardi}, {Collins}, {Conti}, {Dixon},
  {Guerra}, {Haghighipour}, {Herman}, {Hintz}, {Howard}, {Jensen}, {Kielkopf},
  {Kruse}, {Law}, {Martin}, {Maxted}, {Montet}, {Murgas}, {Nelson},
  {Olmschenk}, {Otero}, {Quimby}, {Richmond}, {Schwarz}, {Shporer}, {Stassun},
  {Stephens}, {Triaud}, {Ulowetz}, {Walter}, {Wiley}, {Wood}, {Yenawine},
  {Agol}, {Barclay}, {Beatty}, {Boisse}, {Caldwell}, {Christiansen},
  {Col{\'o}n}, {Deleuil}, {Doyle}, {Fausnaugh}, {F{\H{u}}r{\'e}sz}, {Gilbert},
  {H{\'e}brard}, {James}, {Jenkins}, {Kane}, {Kidwell}, {Kopparapu}, {Li},
  {Lissauer}, {Lund}, {Majewski}, {Mazeh}, {Quinn}, {Quintana}, {Ricker},
  {Rodriguez}, {Rowe}, {Santerne}, {Schlieder}, {Seager}, {Standing},
  {Stevens}, {Ting}, {Vanderspek}, \& {Winn}}]{Kostov2021}
{Kostov}, V.~B., {Powell}, B.~P., {Orosz}, J.~A., {et~al.} 2021, \aj, 162, 234,
  \dodoi{10.3847/1538-3881/ac223a}

\bibitem[{{Kreidberg}(2015)}]{Kreidberg2015}
{Kreidberg}, L. 2015, \pasp, 127, 1161, \dodoi{10.1086/683602}

\bibitem[{Lee \& Chiang(2016)}]{Lee2016}
Lee, E.~J., \& Chiang, E. 2016, The Astrophysical Journal, 817, 90,
  \dodoi{10.3847/0004-637x/817/2/90}

\bibitem[{Li {et~al.}(2016)Li, Holman, \& Tao}]{Li2016}
Li, G., Holman, M.~J., \& Tao, M. 2016, The Astrophysical Journal, 831, 96,
  \dodoi{10.3847/0004-637x/831/1/96}

\bibitem[{Libby-Roberts {et~al.}(2020)Libby-Roberts, Berta-Thompson,
  D{\'{e}}sert, Masuda, Morley, Lopez, Deck, Fabrycky, Fortney, Line,
  Sanchis-Ojeda, \& Winn}]{Libby-Roberts2020}
Libby-Roberts, J.~E., Berta-Thompson, Z.~K., D{\'{e}}sert, J.-M., {et~al.}
  2020, The Astronomical Journal, 159, 57, \dodoi{10.3847/1538-3881/ab5d36}

\bibitem[{Liu {et~al.}(2013)Liu, Zhang, \& Zhou}]{Liu2013}
Liu, H.~G., Zhang, H., \& Zhou, J.~L. 2013, Astrophysical Journal Letters, 767,
  3, \dodoi{10.1088/2041-8205/767/2/L38}

\bibitem[{Martin(2019)}]{Martin2019}
Martin, D.~V. 2019, Monthly Notices of the Royal Astronomical Society, 488,
  3482, \dodoi{10.1093/mnras/stz959}

\bibitem[{Martin \& Triaud(2014)}]{Martin2014}
Martin, D.~V., \& Triaud, A.~H. 2014, Astronomy and Astrophysics, 570,
  \dodoi{10.1051/0004-6361/201323112}

\bibitem[{Martin \& Triaud(2015)}]{Martin2015a}
---. 2015, Monthly Notices of the Royal Astronomical Society, 449, 781,
  \dodoi{10.1093/mnras/stv121}

\bibitem[{Millholland {et~al.}(2020)Millholland, Petigura, \&
  Batygin}]{Millholland2020}
Millholland, S., Petigura, E., \& Batygin, K. 2020, The Astrophysical Journal,
  897, 7, \dodoi{10.3847/1538-4357/ab959c}

\bibitem[{Mills \& Fabrycky(2017)}]{Mills2017}
Mills, S.~M., \& Fabrycky, D.~C. 2017, The Astronomical Journal, 153, 45,
  \dodoi{10.3847/1538-3881/153/1/45}

\bibitem[{Ohno \& Fortney(2022)}]{Ohno2022}
Ohno, K., \& Fortney, J.~J. 2022, The Astrophysical Journal, 930, 50,
  \dodoi{10.3847/1538-4357/ac6029}

\bibitem[{Orosz {et~al.}(2012)Orosz, Welsh, Carter, Brugamyer, Buchhave,
  Cochran, Endl, Ford, MacQueen, Short, Torres, Windmiller, Agol, Barclay,
  Caldwell, Clarke, Doyle, Fabrycky, Geary, Haghighipour, Holman, Ibrahim,
  Jenkins, Kinemuchi, Li, Lissauer, Pr{\v{s}}a, Ragozzine, Shporer, Still, \&
  Wade}]{Orosz2012}
Orosz, J.~A., Welsh, W.~F., Carter, J.~A., {et~al.} 2012, Astrophysical
  Journal, 758, 1, \dodoi{10.1088/0004-637X/758/2/87}

\bibitem[{Orosz {et~al.}(2019)Orosz, Welsh, Haghighipour, Quarles, Short,
  Mills, Satyal, Torres, Agol, Fabrycky, Jontof-Hutter, Windmiller,
  M{\"{u}}ller, Hinse, Cochran, Endl, Ford, Mazeh, \& Lissauer}]{Orosz2019}
Orosz, J.~A., Welsh, W.~F., Haghighipour, N., {et~al.} 2019, The Astronomical
  Journal, 157, 174, \dodoi{10.3847/1538-3881/ab0ca0}

\bibitem[{{Petigura} {et~al.}(2018){Petigura}, {Benneke}, {Batygin}, {Fulton},
  {Werner}, {Krick}, {Gorjian}, {Sinukoff}, {Deck}, {Mills}, \&
  {Deming}}]{Petigura2018}
{Petigura}, E.~A., {Benneke}, B., {Batygin}, K., {et~al.} 2018, \aj, 156, 89,
  \dodoi{10.3847/1538-3881/aaceac}

\bibitem[{Pierens {et~al.}(2020)Pierens, McNally, \& Nelson}]{Pierens2020}
Pierens, A., McNally, C.~P., \& Nelson, R.~P. 2020, Monthly Notices of the
  Royal Astronomical Society, 496, 2849, \dodoi{10.1093/MNRAS/STAA1550}

\bibitem[{Pierens \& Nelson(2013)}]{Pierens2013}
Pierens, A., \& Nelson, R.~P. 2013, Astronomy and Astrophysics, 556, 1,
  \dodoi{10.1051/0004-6361/201321777}

\bibitem[{Pierens {et~al.}(2021)Pierens, Nelson, \& Mcnally}]{Pierens2021}
Pierens, A., Nelson, R.~P., \& Mcnally, C.~P. 2021, Monthly Notices of the
  Royal Astronomical Society, 508, 4806, \dodoi{10.1093/mnras/stab2853}

\bibitem[{Powell {et~al.}(2022)Powell, Kruse, Montet, Feinstein, Lewis,
  Foreman-Mackey, Barclay, Quintana, Colón, Kostov, Boyd, Smale, Mullally,
  Schlieder, Schnittman, Carroll, Carriere, Salmon, Strong, Acks, Pfaff,
  Gerner, \& Burch}]{Powell_2022}
Powell, B.~P., Kruse, E., Montet, B.~T., {et~al.} 2022, Research Notes of the
  AAS, 6, 111, \dodoi{10.3847/2515-5172/ac74c4}

\bibitem[{{Powell, Brian}(2022)}]{https://doi.org/10.17909/j2yt-t417}
{Powell, Brian}. 2022, TESS FFI-Based Light Curves from the GSFC Team
  ("GSFC-ELEANOR-LITE"),  STScI/MAST, \dodoi{10.17909/J2YT-T417}

\bibitem[{{Pulley} {et~al.}(2022){Pulley}, {Sharp}, {Mallett}, \& {von
  Harrach}}]{Pulley2022}
{Pulley}, D., {Sharp}, I.~D., {Mallett}, J., \& {von Harrach}, S. 2022, \mnras,
  514, 5725, \dodoi{10.1093/mnras/stac1676}

\bibitem[{{Qian} {et~al.}(2012){Qian}, {Zhu}, {Dai}, {Fern{\'a}ndez-Laj{\'u}s},
  {Xiang}, \& {He}}]{Qian2012}
{Qian}, S.~B., {Zhu}, L.~Y., {Dai}, Z.~B., {et~al.} 2012, \apjl, 745, L23,
  \dodoi{10.1088/2041-8205/745/2/L23}

\bibitem[{Rappaport {et~al.}(2013)Rappaport, Deck, Levine, Borkovits, Carter,
  {El Mellah}, Sanchis-Ojeda, \& Kalomeni}]{Rappaport2013}
Rappaport, S., Deck, K., Levine, A., {et~al.} 2013, Astrophysical Journal, 768,
  \dodoi{10.1088/0004-637X/768/1/33}

\bibitem[{{Rein} \& {Liu}(2012)}]{Rein&Liu2012}
{Rein}, H., \& {Liu}, S.~F. 2012, \aap, 537, A128,
  \dodoi{10.1051/0004-6361/201118085}

\bibitem[{Rein \& Tamayo(2015)}]{Rein2015}
Rein, H., \& Tamayo, D. 2015, Monthly Notices of the Royal Astronomical
  Society, 452, 376, \dodoi{10.1093/mnras/stv1257}

\bibitem[{Schwamb {et~al.}(2013)Schwamb, Orosz, Carter, Welsh, Fischer, Torres,
  Howard, Crepp, Keel, Lintott, Kaib, Terrell, Gagliano, Jek, Parrish, Smith,
  Lynn, Simpson, Giguere, \& Schawinski}]{Schwamb2013}
Schwamb, M.~E., Orosz, J.~A., Carter, J.~A., {et~al.} 2013, Astrophysical
  Journal, 768, \dodoi{10.1088/0004-637X/768/2/127}

\bibitem[{Sebastian {et~al.}(2024)Sebastian, Triaud, Brogi, Baycroft, Standing,
  Maxted, Martin, Sairam, \& Nielsen}]{Sebastian2024}
Sebastian, D., Triaud, A. H. M.~J., Brogi, M., {et~al.} 2024, 15, 1.
\newblock \doarXiv{2402.06449}

\bibitem[{Socia {et~al.}(2020)Socia, Welsh, Orosz, Cochran, Endl, Quarles,
  Short, Torres, Windmiller, \& Yenawine}]{Socia2020}
Socia, Q.~J., Welsh, W.~F., Orosz, J.~A., {et~al.} 2020, The Astronomical
  Journal, 159, 94, \dodoi{10.3847/1538-3881/ab665b}

\bibitem[{{Standing} {et~al.}(2022){Standing}, {Triaud}, {Faria}, {Martin},
  {Boisse}, {Correia}, {Deleuil}, {Dransfield}, {Gillon}, {H{\'e}brard},
  {Hellier}, {Kunovac}, {Maxted}, {Mardling}, {Santerne}, {Sairam}, \&
  {Udry}}]{Standing2022}
{Standing}, M.~R., {Triaud}, A. H.~M.~J., {Faria}, J.~P., {et~al.} 2022,
  \mnras, 511, 3571, \dodoi{10.1093/mnras/stac113}

\bibitem[{Standing {et~al.}(2023)Standing, Sairam, Martin, Triaud, Correia,
  Coleman, Baycroft, Kunovac, Boisse, Cameron, Dransfield, Faria, Gillon, Hara,
  Hellier, Howard, Lane, Mardling, Maxted, Miller, Nelson, Orosz, Pepe,
  Santerne, Sebastian, Udry, \& Welsh}]{Standing2023}
Standing, M.~R., Sairam, L., Martin, D.~V., {et~al.} 2023, Nature Astronomy, 7,
  702, \dodoi{10.1038/s41550-023-01948-4}

\bibitem[{Sybilski {et~al.}(2013)Sybilski, Konacki, Koz{\l}owski, \&
  He{\l}miniak}]{Sybilski2013}
Sybilski, P., Konacki, M., Koz{\l}owski, S.~K., \& He{\l}miniak, K.~G. 2013,
  Monthly Notices of the Royal Astronomical Society, 431, 2024,
  \dodoi{10.1093/mnras/stt194}

\bibitem[{Team(2021)}]{https://doi.org/10.17909/t9-nmc8-f686}
Team, M. 2021, TESS Light Curves - All Sectors,  STScI/MAST,
  \dodoi{10.17909/T9-NMC8-F686}

\bibitem[{{Ter Braak}(2006)}]{TerBraak2006}
{Ter Braak}, C.~J. 2006, Statistics and Computing, 16, 239,
  \dodoi{10.1007/s11222-006-8769-1}

\bibitem[{{Tokovinin} {et~al.}(2019){Tokovinin}, {Mason}, {Mendez}, {Horch}, \&
  {Brice{\~n}o}}]{Tokovinin2019}
{Tokovinin}, A., {Mason}, B.~D., {Mendez}, R.~A., {Horch}, E.~P., \&
  {Brice{\~n}o}, C. 2019, \aj, 158, 48, \dodoi{10.3847/1538-3881/ab24e4}

\bibitem[{Wang \& Dai(2019)}]{Wang2019}
Wang, L., \& Dai, F. 2019, The Astrophysical Journal, 873, L1,
  \dodoi{10.3847/2041-8213/ab0653}

\bibitem[{Welsh {et~al.}(2012)Welsh, Orosz, Carter, Fabrycky, Ford, Lissauer,
  Pr{\v{s}}a, Quinn, Ragozzine, Short, Torres, Winn, Doyle, Barclay, Batalha,
  Bloemen, Brugamyer, Buchhave, Caldwell, Caldwell, Christiansen, Ciardi,
  Cochran, Endl, Fortney, Gautier, Gilliland, Haas, Hall, Holman, Howard,
  Howell, Isaacson, Jenkins, Klaus, Latham, Li, Marcy, Mazeh, Quintana,
  Robertson, Shporer, Steffen, Windmiller, Koch, \& Borucki}]{Welsh2012}
Welsh, W.~F., Orosz, J.~A., Carter, J.~A., {et~al.} 2012, Nature, 481, 475,
  \dodoi{10.1038/nature10768}

\bibitem[{Welsh {et~al.}(2015)Welsh, Orosz, Short, Cochran, Endl, Brugamyer,
  Haghighipour, Buchhave, Doyle, Fabrycky, Hinse, Kane, Kostov, Mazeh, Mills,
  M{\"{i}}¿½ller, Quarles, Quinn, Ragozzine, Shporer, Steffen, Tal-Or,
  Torres, Windmiller, \& Borucki}]{Welsh2015}
Welsh, W.~F., Orosz, J.~A., Short, D.~R., {et~al.} 2015, Astrophysical Journal,
  809, 26, \dodoi{10.1088/0004-637X/809/1/26}

\bibitem[{{Winn}(2010)}]{winn10}
{Winn}, J.~N. 2010, arXiv e-prints, arXiv:1001.2010.
\newblock \doarXiv{1001.2010}

\bibitem[{Zucker \& Alexander(2009)}]{Zucker2009}
Zucker, S., \& Alexander, T. 2009, Proceedings of the International
  Astronomical Union, 5, 135, \dodoi{10.1017/S1743921309990275}

\end{thebibliography}

\appendix

\section{Inconsistency of binary RV signals between ESPRESSO and HARPS \label{sec:app_harps_residuals}}

In our photodynamical model that jointly fit TESS+ESPRESSO/HARPS data, we found the a peak at binary period still present in HARPS residuals but not in ESPRESSO.

We fit the HARPS+ESPRESSO RV data using the two-Keplerian model using the prescription in \citealt{winn10}. The fitted parameters are semi-amplitude $K$, eccentricity $e$, argument of periastron $\omega$, and time of pericenter passage $T_p$ for each companion. We also include system velocity offset and stellar jitter terms in the parameters lists. We sample the parameter posteriors with MCMC sampling. We initialized 100 walkers and ran 40,000 steps. Convergence is confirmed by ensuring the length of each chain is larger than 50 times auto-correlation times. 
Generally, the fitted parameters and uncertainties are consistent with the result presented in \citetalias{Standing2023}.

\begin{figure}
    \centering
    \includegraphics[width=\textwidth]{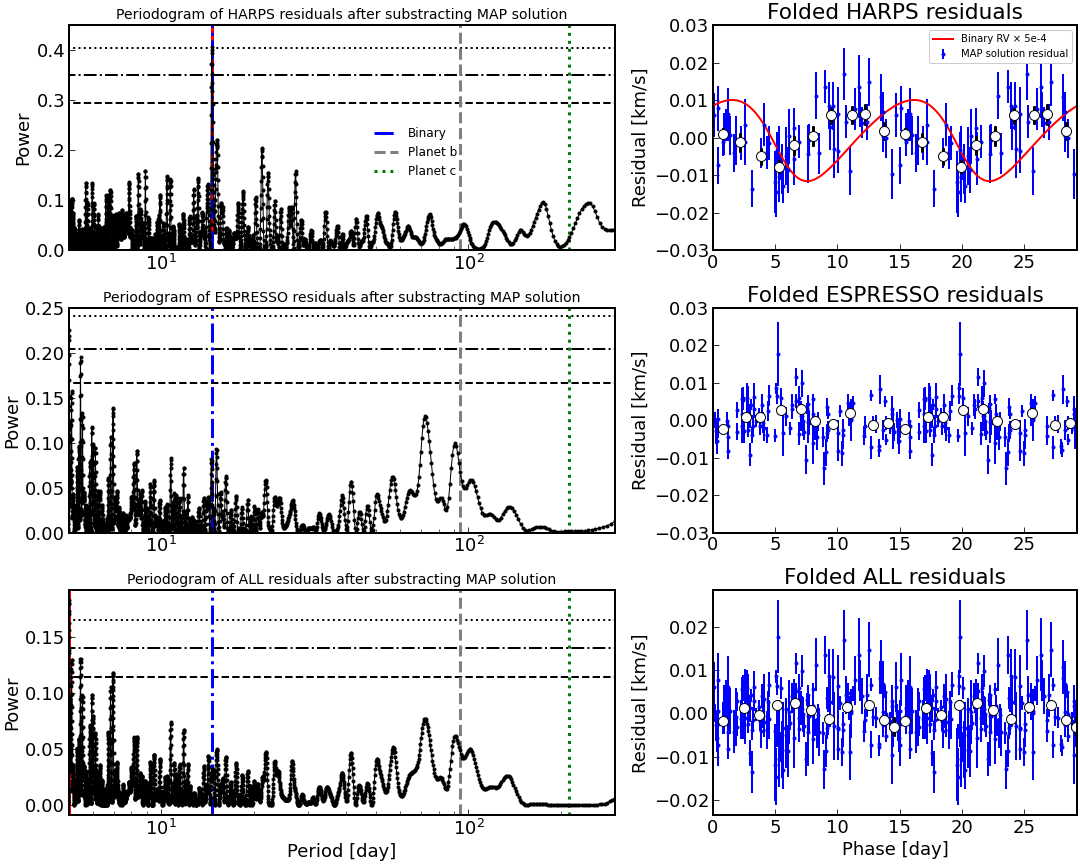}
    \caption{Left: the periodogram of HARPS, ESPRESSO, and all RV observations after subtracting the binary and TOI-1338 c MAP signal from Keplerian orbit fit, the dotted, dot-dashed, and dashed line represent false alarm probability of 0.1\%, 1\%, and 10\% . Right: the phase-folded plot of HARPS ESPRESSO, and all RV residuals at binary period. We show two periods of folded residuals to display the modulation. The binary's Keplerian RV signal is also plotted in red lines to compare with the residual phases.}
    \label{fig:app_harps_residuals}
\end{figure}

Then we subtract the HARPS and ESPRESSO RV data with the MAP solutions and perform a Generalized-Lomb-Scargle periodogram analysis, shown in Figure \ref{fig:app_harps_residuals}. We found no peak at binary periodicity when treating HARPS and ESPRESSO data as a whole or ESPRESSO individually, and the peak binary periodicity is present in HARPS data alone, with a false alarm probability around 0.1\%. The non-detection of binary periodicity in all RV data is probably due to that the number of ESPRESSO data (103) is much larger than the HARPS data (58). 

When we folded the residuals to the binary period, we found a strong modulation in HARPS residuals (top right panel in Figure \ref{fig:app_harps_residuals}), with a peak-to-peak amplitude of $\sim$14 m/s. Compared to the binary RV signal, the residuals excurse the most at the conjunction of the binary, i.e., when the radial velocity is near zero, while in quadrature the excursions are close to zero. 

This inconsistency could be the result of secondary star contamination. As noted in \cite{Hodzic2020}, when the binary is close to conjunction, the radial velocity of primary and secondary components are close and thus the lines are likely blended together, making the shape of CCF change and impact the derived RV from CCF. When the binary is near quadrature, the RVs of both stars are large enough to prevent the overlap of their line cores, a detailed re-analysis of RV measurements that remove this potential cause is beyond the scope of this paper.

\section{Primary/secondary eclipse used in photodynamical modeling \label{sec:pri_sec_eclipse_profile}}

\begin{figure}
    \centering
    \includegraphics[width=0.9\textwidth]{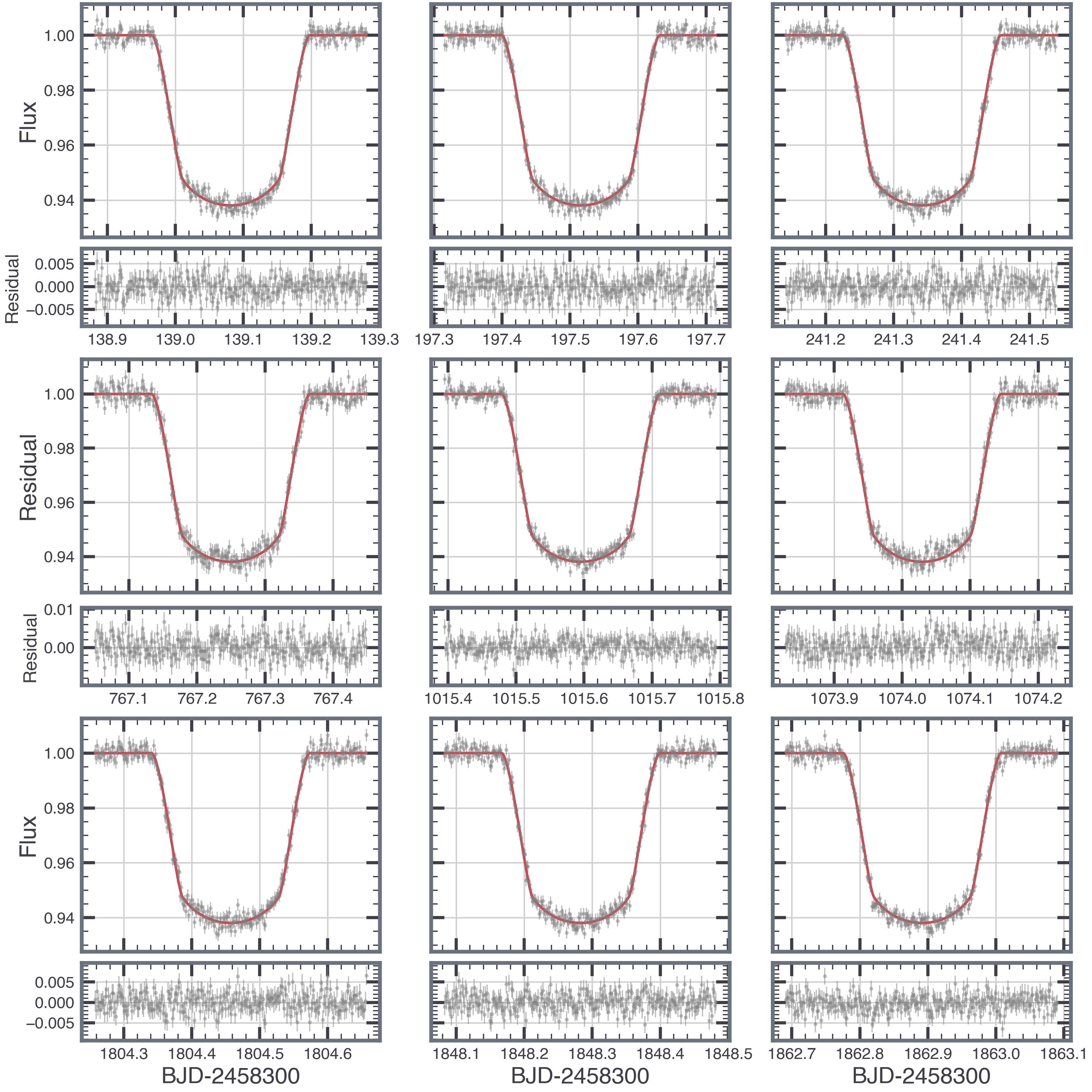}
    \caption{The nine primary eclipse light curves used in the photodynamical modeling. Grey errorbars are TESS 2-min light curves, and red lines are synthetic light curves from the best-fit models. Residuals are plotted below each panel.}
    \label{fig:primary_eclipse_fit}
\end{figure}

\begin{figure}
    \centering
    \includegraphics[width=0.9\textwidth]{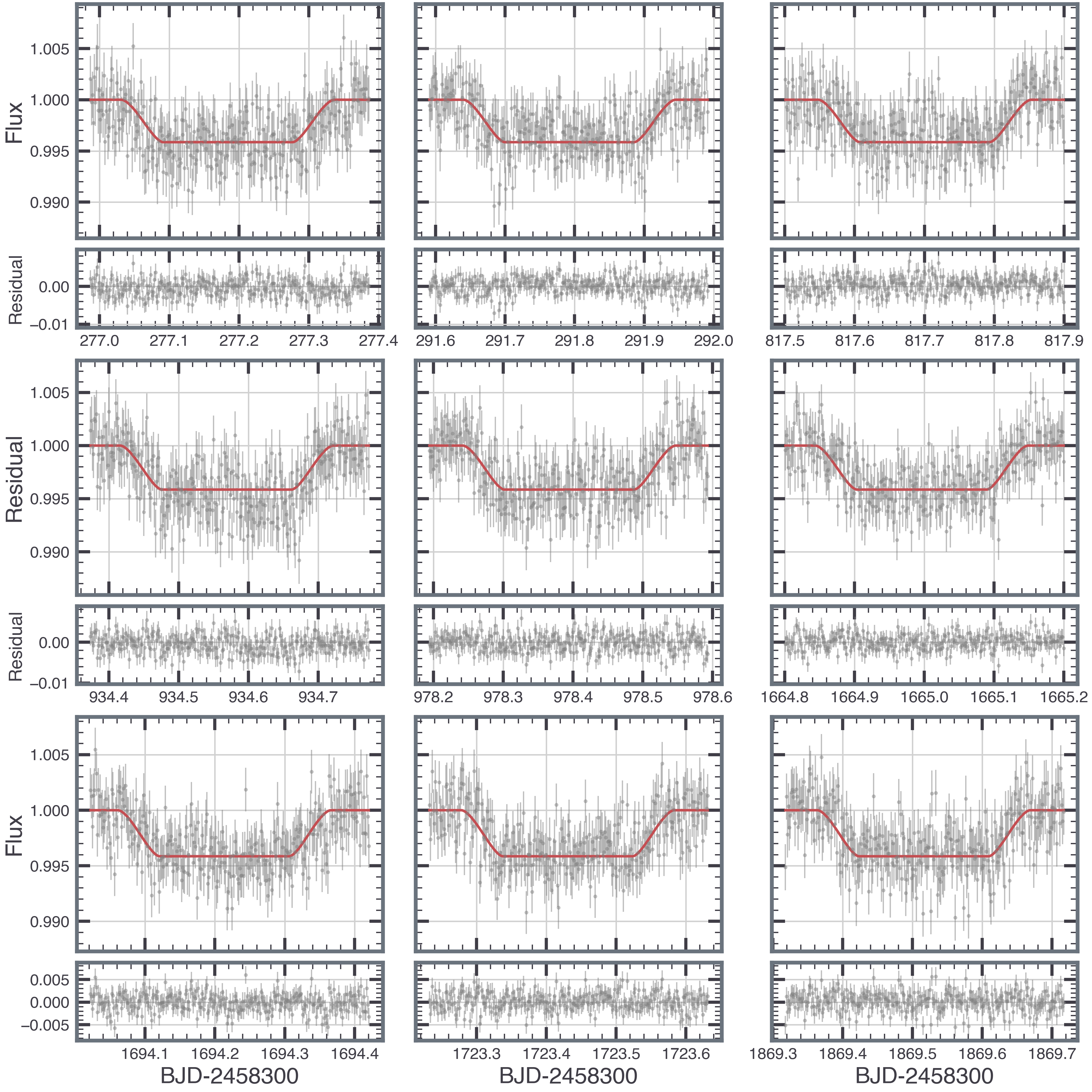}
    \caption{The nine secondary eclipse light curves used in the photodynamical modeling. Labels are the same as Figure \ref{fig:primary_eclipse_fit}}
    \label{fig:secondary_eclipse_fit}
\end{figure}
\end{CJK*}
\end{document}